\documentclass[reqno,10pt,a4paper]{article}
\usepackage{amssymb}
\usepackage{amsmath}
\usepackage{amsthm}
\usepackage{graphicx}
\usepackage[latin1]{inputenc}
\usepackage{epsf,pstricks}

\DeclareMathAlphabet\RsfsCal{U}{rsfs}{m}{n}
\SetMathAlphabet\RsfsCal{bold}{U}{rsfs}{b}{n}

\numberwithin{equation}{section}
\newtheorem{theorem}{Theorem}[section]
     \newtheorem{lemma}[theorem]{Lemma}
     \newtheorem{proposition}[theorem]{Proposition}
     \newtheorem{corollary}[theorem]{Corollary}
     \newenvironment{Proof.}[1][Proof.]{\begin{trivlist}
     \item[\hskip \labelsep {\bfseries #1}]}{\end{trivlist}}

     \newenvironment{acknowledgment}[1][Acknowledgments:]{\begin{trivlist}
     \item[\hskip \labelsep {\bfseries #1}]}{\end{trivlist}}
\theoremstyle{remark}
\newtheorem{remark}[theorem]{Remark}

\newcommand{\field}[1]{\mathbb{#1}}
\newcommand{\C}{\field{C}}
\newcommand{\R}{\field{R}}
\newcommand{\N}{\field{N}}
\newcommand{\overcirc}[1]{{#1}^{{\ }^{{\ }^{\hspace{-4.55 mm} \circ}}}}
\newcommand{\overleq}[1]{{\leq^{^{^{\hspace{-4.7 mm} #1}}}}}

\newcommand{\overeq}[1]{{\; \; =^{^{\hspace{-4.3 mm} #1}}}}

\renewcommand{\Im}{\,\mathrm{Im}\,}
\renewcommand{\Re}{\,\mathrm{Re}\,}

\allowdisplaybreaks

\title{The Cauchy Problem for the Wave Equation in the
Schwarzschild Geometry}
\author{Johann Kronthaler\thanks{Research supported in part by the Deutsche
Forschungsgemeinschaft.}}
\date{January 2006}

\begin{document}
\maketitle

\begin{abstract}
The Cauchy problem is considered for the scalar wave equation in the
Schwarzschild geometry. We derive an integral spectral representation
for the solution and prove pointwise decay in time.
\end{abstract}


\section{Introduction}

Recently pointwise decay was proven for solutions of the scalar wave equation
in the Kerr geometry \cite{Finster1,Finster2}. The main
difficulties in this proof are due to the fact that the metric is
only axisymmetric. In particular, the classical energy density may be
negative inside the \textit{ergosphere}, a region outside the event horizon in
which the Killing vector corresponding to time translations becomes spacelike.
This makes it necessary to apply special methods (spectral theory in Pontrjagin
spaces, energy splitting estimates, causality arguments) which are technically
demanding and not easily accessible. Therefore, it seems worthwile working out
the special case of spherical symmetry (Schwarzschild geometry) separately. This
is precisely the
purpose of the present paper, where we derive an integral representation for
the solution of the Cauchy problem and prove pointwise decay for the scalar
wave equation in the Schwarzschild geometry. In this case, the classical energy
density is positive everywhere outside the event horizon. This gives rise to a
positive definite scalar product, making it possible to apply Hilbert space
methods.

Recall that in Schwarzschild coordinates $(t,r,\vartheta,\varphi)$,
the Schwarzschild metric takes the form
\begin{eqnarray} \label{schwarzschildgeometrie}
ds^2 & \hspace{-3mm} = \hspace{-2mm} & g_{ij} \: dx^{i} dx^{j} \nonumber \\
 & \hspace{-3mm} = \hspace{-2mm} & \left(1- \frac{2M}{r} \right)\: dt^2 -
\left(1-
 \frac{2M}{r}\right)^{-1} dr^2 - r^2(d\vartheta ^2
+ \mathrm{sin}^2 \vartheta \: d\varphi^2)
\end{eqnarray}
with $r>0,\: 0 \leq \vartheta \leq \pi ,\: 0 \leq \varphi < 2\pi$.
We often use for the angular variables the short notation~$x \in S^2$.
Obviously, the metric has two singularities at $r=0$ and $r=2M$. The latter is
called the \textit{event horizon} and can be resolved by a simple coordinate
transformation. In the following we consider only the region $ r> 2M $
outside the event horizon. The scalar wave equation in the Schwarzschild
geometry is given by
\begin{equation} \label{eq: Wellengleichung Grundform}
\square \phi := g^{ij} \nabla_{i} \nabla_{j} \phi =
\frac{1}{\sqrt{-g}} \frac{\partial}{\partial x^{i}} \left( \sqrt{-g} \:g^{ij}
\frac{\partial }{\partial x^{j}} \right) \phi = 0
\end{equation} where $g$ denotes the determinant of the metric $g_{ij}$.
We now state our main result.
\begin{theorem} \label{theorem: Haupttheorem}
Consider the Cauchy problem of the scalar wave equation in the Schwarzschild
geometry
$$ \square \phi = 0\; , \quad (\phi, i \partial_t \phi)(0,r,x) = \Phi_0(r,x)$$
for smooth initial data $\Phi_0 \in C^{\infty}_0 ( (2M,\infty) \times
S^2)^2 $ which is compactly supported outside the event horizon. 
Then there exists a unique global solution \linebreak $\Phi (t) = (\phi (t), i
\partial_t \phi (t)) \in C^\infty(\R \times (2M, \infty) \times S^2)^2$ which is
compactly supported for all times $t$. Moreover, for fixed
$(r,x)$ this solution decays as $t\rightarrow \infty$.
\end{theorem}
There has been considerable work on the wave equation in the Schwarzschild
geometry. In 1957, Regge and Wheeler \cite{ReggeWheeler} investigated the linear
stability of this geometry. Kay and Wald \cite{KayWald} proved
boundedness for solutions of the Klein-Gordon equation in this space-time
outside and on the event horizon. By heuristic arguments, Price \cite{Price}
got evidence for polynomial decay of solutions of the scalar wave equation.
More recently, Dafermos and Rodnianski \cite{Dafermos} gave a mathematical proof
for this decay for spherical symmetric initial data.
For general initial data they derived decay rates~\cite{Dafermos2}, which are
however not sharp.
Furthermore, Morawetz and
Strichartz-type estimates for a massless scalar field without charge in a
Reissner Nordstr{\o}m background with naked singularity are developed in
\cite{Stalker}. And in \cite{Blue} a Morawetz-type inequality was proven for the
semi-linear wave equation in Schwarzschild.

The paper is organized as follows: First, we introduce the Regge-Wheeler
variable and rewrite the wave equation as a
first-order Hamiltonian system. The resulting Hamiltonian is a symmetric
operator
with respect to the scalar product arising from the conserved energy. Exploiting
the spherical symmetry of the problem, we may consider the problem for fixed
angular modes $l$ and $m$. We then show that the corresponding Hamiltonian is
essentially self-adjoint. More precisely, our goal is to apply
Stone's formula, which
relates the propagator to an integral over the resolvent. Thus in Section 4
we give an explicit construction for the resolvent. This
construction is based on special solutions of the radial equation, which
decay
exponentially at $\pm \infty$. In Section 5 we prove the existence
of these solutions via the formalism of the Jost equation. Moreover, we obtain
appropriate regularity results for these solutions, which lead to an integral
representation of the solution operators of the Cauchy problem for fixed $l$
and $m$.
According to the theory of symmetric hyperbolic systems, the Cauchy problem has
a unique smooth solution. Thus, summing over the angular modes yields the
desired representation of this solution. Combining this representation with a
Sobolev imbedding argument, we obtain pointwise decay in time.

\section{Preliminaries}

In this section we reformulate the wave equation as a first order Hamiltonian
system. This will make it possible to analyze the dynamics of the waves with
Hilbert space methods.

According to (\ref{schwarzschildgeometrie}) and (\ref{eq: Wellengleichung
Grundform}) the scalar wave equation in the Schwarzschild geometry with
respect to Schwarzschild coordinates has the explicit form
\begin{equation} \label{Wellengleichung_schwkoord}
\left[ \frac{\partial^2 }{\partial t^2} - \left( 1-\frac{2M}{r}
\right)\:\frac{1}{r^2}\: \left( \frac{\partial}{\partial r} (r^2 - 2Mr)
\frac{\partial}{\partial r} + \Delta_{S^2} \right) \right] \phi =0 \; .
\end{equation}
Here $\Delta_{S^2}$ denotes the standard Laplacian on the two sphere, which
in the coordinates $(\vartheta , \varphi)$ is given by
\begin{equation} \label{eq: Laplace auf der zwei sphaere explizit}
\Delta_{S^2} =  \frac{1}{\sin^2 \vartheta} \frac{\partial^2}{\partial \varphi
^2}+ \frac{\partial}{\partial (\cos \vartheta)} \sin^2 \vartheta
\frac{\partial}{\partial (\cos \vartheta)} \; .
\end{equation} 
In order to bring the equation (\ref{Wellengleichung_schwkoord}) into a more
convenient form, we first introduce the Regge-Wheeler coordinate $u$ by
\begin{equation} \label{reggewheelercoord}
u(r) := r + 2M\: \mathrm{log} \left(\frac{r}{2M} -1 \right) \: .
\end{equation}
The variable $u$ takes values in the whole interval $(-\infty,\infty)$ as $r$
ranges over $(2M,\infty)$. It satisfies the relations
\begin{equation} \label{ableitung r w coord}
\frac{du}{dr} = \frac{1}{1 - \frac{2M}{r}} \:, \qquad \frac{\partial}{\partial
u} =
\left( 1 - \frac{2M}{r} \right) \frac{\partial}{\partial r} \: .
\end{equation} In what follows the variable $r$ is always implicitly
given by $u$. Using the Regge-Wheeler coordinate, the wave equation
(\ref{Wellengleichung_schwkoord}) transforms to
\begin{equation} \label{Wellengleichung r w coord}
\left[ \frac{ \partial^2}{\partial t^2}   - \frac{1}{r} \:
\frac{\partial^2}{\partial u^2} \: r + \left( 1- \frac{2M}{r} \right) \left(
\frac{2M}{r^3} - \frac{\Delta_{S^2}}{r^2}  \right) \right] \phi = 0 \: .
\end{equation}
To simplify this equation we multiply by $r$ and substitute
$\phi =  r \psi$ This leads us to the Cauchy problem
\begin{equation} \label{eq: Cauchy problem in 4 Dimensionen}
\left. \begin{array}{c}
\displaystyle \left[ \frac{ \partial^2}{\partial t^2} -
\frac{\partial^2}{\partial u^2} +
\left( 1- \frac{2M}{r} \right) \left(
\frac{2M}{r^3} - \frac{\Delta_{S^2}}{r^2}  \right) \right]
\psi(t,u,x) = 0 \\
\ \\
(\psi, i \partial_t \psi)(0,u,x) = \Psi_0 (u,x)
\end{array} \right\}
\end{equation}
where the initial data $\Psi_0 \in C_0^\infty(\R \times S^2)^2$ is smooth and
compactly supported.

\bigskip

The equation in (\ref{eq: Cauchy problem in 4 Dimensionen}) can be reformulated
as the Euler-Lagrange equation corresponding to the action
\begin{equation} \label{action}
S = \int_{-\infty}^{\infty} dt \int_{-\infty}^{\infty} du \int_{-1}^1 d(\cos
\vartheta) \int_0^{2\pi} d\varphi \: \mathcal{L}(\psi,
\nabla \psi) \: ,
\end{equation} 
where the Lagrangian is given by
\begin{eqnarray} 
2 \: \mathcal{L} & = & | \partial_t \psi |^2 - | \partial_u \psi |^2 -
\left(1- \frac{2M}{r}\right) \frac{2M}{r^3} |\psi|^2 - \nonumber \\
& & \left(1- \frac{2M}{r}\right)\frac{1}{r^2} \left( \frac{1}{\sin^2 \vartheta}
|\partial_\varphi \psi|^2 + \sin^2\vartheta \: |\partial_{\cos \vartheta} \:
\psi|^2\right)  \; . \label{Lagrangian}
\end{eqnarray}
As one sees immediately, the Lagrangian is invariant under time translations,
and thus Noether's theorem gives rise to a conserved quantity, the
energy $E$,
\begin{equation} \label{energy}
E[\psi] = \int_{-\infty}^\infty du \int_{-1}^1 d(\cos \vartheta) \int_0^{2 \pi}
\frac{d \varphi}{\pi} \: \mathcal{E} ,
\end{equation}
where $\mathcal{E}$ is the energy density
\begin{equation*} 
2 \mathcal{E}  =  2 \left( \frac{\partial \mathcal{L}}{\partial \psi_t} \:
\psi_t -\mathcal{L} \right)  =  | \partial_t \psi |^2 + |
\partial_u \psi |^2+ \hspace{50mm}
\end{equation*}
\begin{equation} \label{energy density}
 +\left(1- \frac{2M}{r}\right) \bigg\{ \frac{2M}{r^3} |\psi|^2 
+ \frac{1}{r^2} \left( \frac{1}{\sin^2 \vartheta}
|\partial_\varphi \psi|^2 + \sin^2\vartheta \: |\partial_{\cos \vartheta} \:
\psi|^2\right) \bigg\} \; .
\end{equation} 
It is also easy to check directly that the above energy is conserved
in time for all smooth solutions of the wave equation
that are compactly supported for all times. Since we consider the wave equation
outside the event horizon, i.e. $r>2M$, it is clear that the energy density is
positive everywhere.

\bigskip

Next we rewrite the Cauchy problem (\ref{eq: Cauchy problem in 4 Dimensionen})
in first-order Hamiltonian form. \\
Letting
\begin{equation} \label{phi dtphi}
\Psi = \left(%
\begin{array}{c}
 \! \! \psi \! \! \\
\! \! i \partial_t \psi \! \! \\
\end{array}%
\right) \; ,
\end{equation}
the Cauchy problem takes the form
\begin{equation} \label{Hamiltonform allgemein}
i \partial_t \Psi = H \Psi \; , \quad \Psi \big|_{t=0} = \Psi_0
\end{equation}
where $H$ is the Hamiltonian
\begin{equation} \label{Hamiltonian allgemein}
\left(%
\begin{array}{cr}
  0 & 1 \\
 A & 0 \\
\end{array}%
\right) \; .
\end{equation}
Here $A$ is the differential operator
\begin{equation} \label{eq: Differentialoperator A}
A = - \partial_u^2 +\left( 1-
\frac{2M}{r} \right) \left(
\frac{2M}{r^3} - \frac{1}{r^2}\Delta_{S^2}  \right) \; .
\end{equation}
We use the energy $E$ in order to introduce a scalar product such that the
Hamiltonian $H$ is symmetric with respect to it. More precisely, we endow the
space $C_0^\infty (\R \times S^2)^2$ with the \textit{energy scalar product}
$\langle . , .\rangle$ by polarizing $E$, thus
\begin{equation*} 
 \langle \Psi , \Phi \rangle  := \int_{-\infty}^\infty du
\int_{-1}^1 d(\cos \vartheta ) \int_0^{2\pi}\frac{d \varphi}{2 \pi} \bigg\{
\overline{\partial_t \psi} \partial_t \phi + \overline{\partial_u \psi}
\partial_u \phi +  \left(1- \frac{2M}{r}\right) \hspace{10mm}
\end{equation*}
\begin{equation} \label{definition energy scalar product}
\hspace{11mm} \times \left[ \frac{2M}{r^3} \:
\overline{\psi} \: \phi + \frac{1}{r^2} \left( \frac{1}{\sin^2 \vartheta} \:
\overline{\partial_\varphi \psi} \partial_\varphi \phi + \sin^2\vartheta \:
\overline{\partial_{\cos \vartheta} \: \psi} \partial_{\cos \vartheta} \:
\phi \right) \right]
 \bigg\}\;  ,
\end{equation}
where again $\Psi =  (\psi,  i \partial_t \psi)^T$ and $\Phi = (\phi, i
\partial_t \phi)^T$. Energy conservation implies that for a solution $\Psi$ of
the Cauchy problem (\ref{Hamiltonform allgemein}) which is compactly supported
for all times,
\begin{eqnarray*}
0 & = & \frac{d}{dt} E[\Psi] = \frac{d}{dt} \langle \Psi , \Psi \rangle =   \\
\\
 & = & \langle \dot{\Psi} , \Psi \rangle + \langle \Psi , \dot{\Psi} \rangle
= i\langle H \Psi, \Psi \rangle - i \langle \Psi, H \Psi \rangle \; .
\end{eqnarray*}
Since the initial data $\Psi_0 \in C^\infty_0 (\R \times S^2)^2$ can be chosen
arbitrarily, polarization yields
\begin{equation} \label{Hsymmetrisch}
\langle H\Psi, \Phi \rangle = \langle \Psi , H \Phi \rangle\: , \quad
\textrm{for all } \Psi,\Phi \in C^{\infty}_0(\mathbb{R} \times S^2)^2 \; .
\end{equation}
Hence the operator $H$ is symmetric on $C_0^\infty(\R \times S^2)^2$ with
respect to $\langle . , . \rangle $.

We will now use the spherical symmetry to simplify the problem. More
precisely, we make use of the fact that the angular dependence of the
wave equation in the Schwarzschild geometry involves only the Laplacian on the
two sphere. It is well-known that the spherical harmonics $ \{ Y_{lm}(\vartheta,
\varphi) \}_{l \in \N_0, |m| \leq l}$ are smooth eigenfunctions of
$\Delta_{S^2}$ with the eigenvalues $ -l(l+1)$. Moreover, they form an
orthonormal basis of the space $L^2 (S^2)$. Thus we can decompose an arbitrary
$\Psi \equiv
(\psi_1,\psi_2)^T \in C^\infty_0 (\R \times S^2)^2$ in the following way,
\begin{equation} \label{eq: Dekomposition von Psi in spherical harmonics}
\Psi(u,\vartheta, \varphi) =  \sum_{l=0}^\infty \sum_{|m| \leq l} \Psi^{lm}(u)
Y_{lm}( \vartheta, \varphi) \; ,
\end{equation}
where for each component the sum converges for fixed $u$ in $L^2 (S^2)$. Since
the $\Psi^{lm} \equiv (\psi^{lm}_1, \psi^{lm}_2)^T$ are uniquely determined by
$\psi^{lm}_i(u) = \langle Y_{lm} , \psi_i(u)\rangle_{L^2(S^2)}$ it is clear that
$\Psi^{lm}(u) \in C^\infty_0(\R)^2$ for all $l,m$. Using this decomposition,
we rewrite the norm of $\Psi$ corresponding to the energy scalar product as
\begin{eqnarray}
\langle \Psi, \Psi \rangle =  \int_{-\infty}^\infty \!du \!
\int_1^1
\! d(\cos \vartheta) \! \int_0^{2\pi} \! \frac{d \varphi}{2\pi} \: \bigg\{
| \psi_2 |^2   + |\partial_u \psi_1|^2 +\hspace{25mm} \nonumber \\ 
\nonumber  + \overline{\psi_1} \left (1
- \frac{2M}{r}\right) \left( \frac{2M}{r^3} - \frac{1}{r^2} \Delta_{S^2} \right)
\psi_1 \bigg\}  \\
 \nonumber = \sum_{l=0}^\infty \sum_{|m| \leq l} \int_{-\infty}^\infty \! du
\bigg \{ | \psi^{lm}_2(u)|^2 + | \partial_u \psi^{lm}_1(u)|^2 + \hspace{27mm} \\
\label{eq: Darstellung Energieskalarprodukt ueber Summe der Moden}
\left (1 -
\frac{2M}{r}\right) \left( \frac{2M}{r^3} + \frac{l(l+1)}{r^2} \right)
|\psi^{lm}_1(u)|^2 \bigg\}  \; , \hspace{5mm}
\end{eqnarray}
where in the first equation we have integrated by parts with respect
to $(\vartheta , \varphi)$. The second equation follows from the
properties of the $Y_{lm}$. As one can immediately see, the integrand for every
summand in (\ref{eq: Darstellung Energieskalarprodukt ueber Summe der Moden}) is
positive. Hence again by
polarizing we obtain for any angular mode $l$ an scalar product $\langle . ,.
\rangle _l$ on $C^\infty_0(\R)^2$ given by
\begin{equation} \label{eq: Energieskalarprodukt fuer die mode l}
\langle \Psi , \Phi \rangle_l = \int_{-\infty}^\infty \! \Big \{ 
\overline{\psi_2} \phi_2  + \overline{\psi'_1} \phi'_1 + V_l\:
\overline{\psi_1} \phi_1 \Big\} \:du   \; ,
\end{equation} 
with the potential $V_l(u)$ defined as
\begin{equation} \label{potential}
V_l(u) = \left( 1- \frac{2M}{r} \right) \left(\frac{2M}{r^3} +
\frac{l(l+1)}{r^2}
\right) \: .
\end{equation}
This definition leads to an isometry
\begin{eqnarray}
\big( C^\infty_0 (\R \times S^2)^2  , \langle . , . \rangle \big)
& \longrightarrow & \bigoplus_{l= 0}^\infty \bigoplus_{|m| \leq l}
\big(C^\infty_0
(\R)^2 , \langle . , . \rangle_l \big) \nonumber \\
 \Psi & \mapsto & \Psi^{lm} \; . 
\label{eq: Isometrie bzgl der Zerlegung in Moden}
\end{eqnarray} 
Using (\ref{eq: Dekomposition von Psi in spherical harmonics}), the Hamiltonian
$H$ also decomposes in the following way,
$$ H \Psi(u,\vartheta, \varphi) =  \sum_{l=0}^\infty \sum_{|m| \leq l}
H_l \Psi^{lm}(u) Y_{lm}( \vartheta, \varphi) \; .$$
Here the $H_l$ act on $C^\infty_0(\R)^2$ and are given by
\begin{equation} \label{Hamiltonian}
H_l = \left( \begin{array}{cr}
0 &  1\\ 
-\partial_u^2 + V_l(u)  &  0 
\end{array} \right) \; .
\end{equation}
Thus for fixed angular modes $l$ and $m$ the Cauchy problem (\ref{Hamiltonform
allgemein}) simplifies to
\begin{equation} \label{Hamiltonform}
i \partial_t \Psi^{lm} =  H_l \Psi^{lm} \; ,\quad \Psi^{lm}\big|_{t= 0} =
\Psi_0^{lm} \; ,
\end{equation}
where the initial data is in $C^\infty_0(\R)^2$. 
Moreover, the $H_l$ are symmetric on $C_0^\infty(\R)^2$ with respect to $\langle
. ,
.\rangle_l$, because for any  $\Psi,\Phi \in C^\infty_0(\R)^2$ the
functions $ \Psi(u) Y_{lm}$ and $ \Phi(u) Y_{lm}$ are in $C^\infty_0 (\R \times
S^2)^2$. Thus
$$ \langle H_l \Psi ,\Phi \rangle_l = \langle H (\Psi \: Y_{lm}) , \Phi Y_{lm}
\rangle = \langle  \Psi Y_{lm} , H (\Phi \:  Y_{lm}) \rangle = \langle  \Psi ,
H_l \Phi \rangle_l \; .$$ In particular, for solutions of
(\ref{Hamiltonform}) with compact support in $u$ for all times, the norm with
respect to $\langle. ,. \rangle_l$ is constant. Therefore we again refer to
$\langle . , .\rangle_l$ as the energy scalar product.

Our strategy is to solve for a given inital data $ \Psi_0 \in C^\infty_0(\R
\times S^2)^2$ the Cauchy problem (\ref{Hamiltonform}) for fixed angular modes
$l$ and $m$, and to sum up the solutions afterwards. Therefore, in what follows
we will
fix the angular modes $l,m$ and consider the problem (\ref{Hamiltonform}). In
order to avoid too many indices, we usually omit the subscript $l$ in the
Hamiltonian and
energy scalar product.

\section{Spectral Properties of the Hamiltonian}

In the previous section we introduced the energy scalar product $\langle \: ,
\rangle$ on the space $ C^{\infty}_0(\mathbb{R})^2$. Since we cannot expect $
C^{\infty}_0(\mathbb{R})^2$ to be complete with respect to this inner product
(and indeed it is not, because the energy scalar product in the second component
is just the usual $L^2$-scalar product), we define the Hilbert space $H^1_{V_l
0}(\mathbb{R})$ as the completion of $C^{\infty}_0(\mathbb{R})$ within the 
Hilbert space
$$ H^1_{V_l}(\mathbb{R}) = \left \{ u  \: \textrm{with} \: u' \in
L^2(\mathbb{R}) \textrm{ and }
V^{1/2}_l u \in L^2(\mathbb{R}) \right \}$$ endowed with the scalar product
$$  \langle u ,v \rangle_1 := (u',v')_{L^2} + ( V_l \: u, v)_{L^2} \; .$$
Note that this coincides with the energy scalar product on the first
component. Therefore, we choose  $ \algcal{H} \equiv H^1_{V_l  0}(\mathbb{R})
\oplus L^2(\mathbb{R}) $ endowed with the energy scalar product as the
underlying Hilbert space for our Hamiltonian $H$.

\bigskip

In the previous section we have seen that the Hamiltonian $H$ is symmetric on
$C_0^\infty(\R)^2$. Before we can use functional analytic methods, we need to
construct a
self-adjoint extension of $H$. In fact, we are able to prove the following
lemma:

\begin{lemma} \label{H ess s.a.}
The operator $H$ with domain $\algcal{D}(H) = C_0^\infty(\mathbb{R})^2$ is
essentially self-adjoint in the Hilbert space $\algcal{H}$.
\end{lemma}

In order to prove this lemma, we use the following version of Stone's theorem
about strongly continuous one-parameter unitary groups. A proof of this theorem
can be found in \cite[Section VIII.4]{RS1}.

\begin{theorem} \label{stones theorem}
Let $U(t)$ be a strongly continuous one-parameter unitary group on a Hilbert
space $\algcal{H}$. Then there is a self-adjoint operator $A$ on $\algcal{H}$
such that $U(t) = e^{itA}$.

Furthermore, let $D$ be a dense domain which is invariant under $U(t)$ and on
which $U(t)$ is strongly differentiable. Then $i^{-1}$ times the strong
derivative of $U(t)$ is essentially self-adjoint on $D$, and its closure is $A$.
\end{theorem} Now we apply this theorem:

\begin{proof}[Proof of Lemma \ref{H ess s.a.}]
According to the theory of symmetric hyperbolic systems (cf. \cite[Section
5.3]{John}), the Cauchy problem 
$$ \left. \begin{array}{c} \left(\partial_t^2 - \partial_u^2 + V_l(u)\right)
\psi(t,u) = 0 \\ \vspace{-2mm} \\ \psi|_{t=0} = f \; , \: i\partial_t \psi
|_{t=0} = g \end{array} \right\}$$ with smooth, compactly
supported initial data $f,g \in
C_0^\infty(\mathbb{R})$ has a unique solution $\psi(t,u) \in
C^\infty(\mathbb{R} \times \mathbb{R})$ which is also compactly supported in
$u$ for all times.
Using this solution, we define for arbitrary $t \in \mathbb{R}$ the operators
\begin{eqnarray*}
U(t):  C_0^{\infty}(\mathbb{R})^2 & \rightarrow & C_0^{\infty}(\mathbb{R})^2 \\
[-3mm]\\
\left(%
\begin{array}{c}
  f \\
  g \\
\end{array}%
\right) & \mapsto & \left(%
\begin{array}{c}
  \psi(t,.) \\
 i \partial_t \psi (t,.) \\
\end{array}%
\right) \; ,
\end{eqnarray*}
which leave the dense subspace $C_0^{\infty}(\mathbb{R})^2 \subseteq \algcal{H}$
invariant for all times $t$. \\
Due to the energy conservation, the $U(t)$ are
unitary with respect to the energy scalar product and hence extend to unitary
operators
on the entire Hilbert space $\algcal{H}$. Furthermore, since the solution is
uniquely determined by the initial data, the $U(t)$ have the following
properties,
$$ U(0) = Id, \quad U(t+s) = U(t)U(s) \quad \textrm{for all } t,s \in \mathbb{R}
\; ,$$
and thus they form a one-parameter unitary group. Due to the fact
that smooth initial data yields smooth solutions in $t$ and $u$, this group
is strongly continuous on $\algcal{H}$ and strongly differentiable on the
domain $C_0^\infty(\mathbb{R})^2$. Calculating $i^{-1}$ times the strong
derivative one gets
$$ i^{-1} \lim_{h \searrow 0} \frac{1}{h}\left(U(h)\left(%
\begin{array}{c}
  f \\
  g \\
\end{array}%
\right) - \left(%
\begin{array}{c}
  f \\
  g \\
\end{array}%
\right) \right) = i^{-1} \left(%
\begin{array}{c}
  -i g \\
  i (\partial_u ^2 - V_l) f  \\
\end{array}%
\right) = - H \left(%
\begin{array}{c}
  f \\
  g \\
\end{array}%
\right)$$ for all $ f,g \in C^{\infty}_0(\mathbb{R}),$ and the lemma follows
from Theorem \ref{stones theorem}.
\end{proof}

\bigskip

For the further investigations of the Hamiltonian $H$, we consider its
self-adjoint closure which, for the sake of simplicity, we again denote by
$H$. For our purposes, it is not important to know the exact domain of
definition $\algcal{D} (H)$ of the self-adjoint extension.
\bigskip

\section{Construction of the Resolvent} \label{section: construction of the
resolvent}

Stone's formula for the spectral projections of a self-adjoint operator
$A$ (cf. \cite{RS1} Theorem VII.13),
\begin{equation} \label{stones formula}
\frac{1}{2} \left[ P_{[a,b]} + P_{(a,b)} \right]
=\textrm{s-}\hspace{-1mm}\lim_{\epsilon \searrow 0} \frac{1}{2\pi i} \int_a^b
\left[(A- \omega - i\epsilon)^{-1} - (A- \omega + i \epsilon)^{-1} \right] \: d
\omega \; ,
\end{equation} 
relates the spectral projections to the resolvent (here s-$\lim$ denotes the
strong limit of operators). In view of this relation, it is of interest to
derive an explicit representation of the resolvent.

\bigskip

In the preceding section we have seen that there is a domain $\algcal{D}(H) $
such that our Hamiltonian $H$ is self-adjoint in the Hilbert space $(\algcal{H}
,\langle \: , \rangle)$. From this it immediately follows that the spectrum
$\sigma(H) \subseteq \mathbb{R}$ is on the real line and therefore the
resolvent $(H -\omega)^{-1} : \algcal{H} \rightarrow \algcal{H}$ exists for
every $\omega \in \mathbb{C} \setminus \mathbb{R}$.

\bigskip

Let us now fix $\omega \in \C \setminus \R$. We often denote the
$\omega$-dependence by a subscript $_{\omega}$ .We begin by reducing the
eigenvalue equation $ H \Psi = \omega \Psi$ by substituting the equation for the
first component in the second equation. We thus obtain the
Schr\"odinger-type equation
\begin{equation} \label{Schroedinger equation}
\left( -\partial_u^2  +V_\omega(u) \right) \phi(u) = 0
\end{equation}
with the potential
\begin{equation} \label{schroedinger potential}
V_\omega(u) = - \omega^2 + V_l(u) = - \omega^2 + \left( 1- \frac{2M}{r} \right)
\left(\frac{2M}{r^3} + \frac{l(l+1)}{r^2} \right) \; .
\end{equation}

In what follows we refer to this equation simply as the Schr\"odinger
equation. It can be regarded as the radial equation associated to the wave
equation in (\ref{eq: Cauchy problem in 4 Dimensionen}). Our goal is to
construct the resolvent $(H - \omega)^{-1}$ out of special solutions of this
equation. We introduce fundamental solutions $\acute{\phi}_\omega$
and $\grave{\phi}_{\omega}$ of the Schr\"odinger equation (\ref{Schroedinger
equation}) which satisfy asymptotic boundary conditions at $u= \pm \infty$ (the
existence of these solutions will be proved in Section 5). More precisely, in
the case $\Im (\omega) > 0 $ we impose that
\begin{eqnarray}
 \lim_{u \rightarrow -\infty}  e^{i \omega u} \acute{\phi}_{\omega}(u) =1
\; ,  \quad  & \displaystyle \hspace{4mm} \lim_{u \rightarrow -\infty}  \left(
e^{i \omega u} \acute{\phi}_{\omega}(u) \right)' =
0 \textrm{\ \ \ } \label{boundarycond1} \\
\lim_{u \rightarrow +\infty}  e^{-i \omega u} \grave{\phi}_{\omega}(u) =1 \; ,
\quad & \displaystyle  \lim_{u \rightarrow +\infty}  \left( e^{-i \omega u}
\grave{\phi}_{\omega}(u) \right)' = 0 \; ,\label{boundarycond2}
\end{eqnarray}
whereas in the case $ \Im(\omega) < 0$,
\begin{eqnarray}
\lim_{u \rightarrow -\infty}  e^{-i \omega u} \acute{\phi}_{\omega}(u) =1 \; ,
\quad & \displaystyle \lim_{u \rightarrow -\infty}  \left( e^{-i \omega u}
\acute{\phi}_{\omega}(u) \right)' =
0  \label{boundarycond3} \\
\lim_{u \rightarrow +\infty}  e^{i \omega u} \grave{\phi}_{\omega}(u) =1 \; ,
\quad & \displaystyle \hspace{4mm} \lim_{u \rightarrow +\infty}  \left( e^{i
\omega u}
\grave{\phi}_{\omega}(u) \right)' = 0 \; . \label{boundarycond4}
\end{eqnarray}

Since the resolvent exists, the map $(H-\omega): \algcal{D}(H) \rightarrow
\algcal{H}$ is bijective and in particular the kernel is trivial. Hence the
solutions $ \acute{\phi}_{\omega}, \grave{\phi}_{\omega}$ are linearly
independent (otherwise they would give rise to a vector in the kernel due to
the exponential decay). Thus $\acute{\phi}_\omega$ and $\grave{\phi}_\omega$
are indeed a system of fundamental solutions with non-vanishing Wronskian
\begin{equation} \label{wronski}
w(\acute{\phi}_{\omega},\grave{\phi}_{\omega}) := \acute{\phi}_{\omega}(u)
\grave{\phi}_{\omega}'(u) - \acute{\phi}_{\omega}'(u) \grave{\phi}_{\omega}(u)
\; .
\end{equation}
Note that the Wronskian is independent of the variable $u$, as is easily
verified by differentiating with respect to $u$ and substituting the
Schr\"odinger equation.

\bigskip

In the next lemma, we use this fundamental system to derive the Green's
function corresponding to (\ref{Schroedinger equation}).
\begin{lemma} \label{lemma zu function s(u,v)}
The function
\begin{equation} \label{s(u,v)}
s_\omega(u,v) := - \frac{1}{w(\acute{\phi}_{\omega},\grave{\phi}_{\omega})}
\cdot
\left\{
\begin{array}{cl}
\acute{\phi}_{\omega}(u) \grave{\phi}_{\omega}(v)\:, & \textrm{if } u \leq v \\
\acute{\phi}_{\omega}(v) \grave{\phi}_{\omega}(u) \: , & \textrm{if } u > v
\end{array} \right.
\end{equation}
satisfies the distributional equations
$$ \left( - \frac{\partial^2}{\partial u^2} + V_\omega(u) \right) s_\omega(u,v)
=
\delta(u-v)= \left( - \frac{\partial^2}{\partial v^2} + V_\omega(v) \right)
s_\omega(u,v) \; .$$
\end{lemma}

\vspace{0.1 cm}

\begin{proof}
By definition of the distributional derivative we have for every test
\linebreak function $\eta
\in C^\infty_0(\mathbb{R})$,
$$ \int_{-\infty}^\infty \eta (u)\left[ - \partial_u ^2 + V_\omega(u)\right]
s_\omega(u,v)\:du = \int_{-\infty}^\infty \left[ ( - \partial_u ^2 +
V_\omega(u) ) \eta (u) \right] s(u,v) \: du \; . $$
It is obvious from its definition that the function $s(.,v)$ is smooth except at
the point $u =v$, where its first derivative has a discontinuity. Thus, after
splitting up the integral, we can integrate by parts twice to obtain
\begin{eqnarray*}
\int_{-\infty}^\infty \left[ ( - \partial_u ^2 + V_\omega(u) ) \eta (u) \right]
s(u,v) \: du = \qquad \qquad \qquad \qquad \qquad \\
= \int_{-\infty}^v  \eta (u) ( - \partial_u ^2 + V_\omega(u) )
s(u,v) \:du + \lim_{u \nearrow v} \left[ \eta (u) \partial_u s(u,v) \right]+ \\
+ \int_v^\infty   \eta (u) ( - \partial_u ^2 + V_\omega(u) ) s(u,v) \: du -
\lim_{u \searrow v}\left[ \eta (u) \partial_u s(u,v)\right] \; .
\end{eqnarray*}
Since for $u \neq v$, $s$ is a solution of (\ref{Schroedinger equation}), the
obtained integrals vanish. Computing the limits with the definition
(\ref{s(u,v)}), we get
\begin{eqnarray*}
\int_{-\infty}^\infty \left[ ( - \partial_u ^2 + V_\omega(u) ) \eta (u) \right]
s(u,v) \: du = \left( \lim_{u \nearrow v} - \lim_{u \searrow v} \right) \eta
(u) \partial_u s(u,v) =\\
= - \frac{1}{w(\acute{\phi}_{\omega},\grave{\phi}_{\omega})} \: \eta(v) \left[
\acute{\phi}'_\omega(v) \grave{\phi}_\omega (v) - \grave{\phi}'_\omega (v)
\acute{\phi}_\omega (v) \right] = \eta(v) \; ,
\end{eqnarray*}
where in the last step we used the definition of the Wronskian (\ref{wronski}).
This yields the first equation. The second equation is proven exactly in the
same way.
\end{proof}

With this function $s$ we are now able to construct the resolvent. More
\linebreak precisely,

\begin{proposition} \label{construction of the resolvent}
For every $\omega \in \C \setminus \R$, the resolvent $(H - \omega)^{-1} :
\algcal{H} \rightarrow \algcal{H}$ can be represented as an integral operator
with the integral kernel
\begin{equation} \label{integral kernel k}
k_\omega (u,v) = \delta(u-v) \left(%
\begin{array}{cc}
  0 & 0 \\
  1 & 0 \\
\end{array}%
\right) + s_\omega (u,v) \left(%
\begin{array}{lc}
  \omega & 1 \\
  \omega^2 & \omega \\
\end{array}%
\right) \; .
\end{equation}
\end{proposition}

\begin{proof}
We introduce the integral operator $S_\omega$ with the integral kernel
$k_\omega (u,v) $ on the domain
$$\algcal{D}(S_\omega) := \left \{ (H- \omega) \Psi  \: \big| \:
 \Psi \in C^\infty_0(\mathbb{R})^2  \right\}
\; .$$
Let us verify that $\algcal{D}(S_\omega)$ is a dense subset of $\algcal{H}$. Let
$\phi \in \algcal{H}$ be an arbitrary vector. Because of the existence
of the resolvent, the operator $H- \omega: \algcal{D}(H) \rightarrow \algcal{H}$
is onto, and thus there is a vector $\psi \in \algcal{D}(H)$ with $(H -\omega)
\psi = \phi$. Then due to the definition of the closure of~$H$,
there is a sequence $ \{ \psi_n \}_{n \in \mathbb{N}}
\subseteq C^\infty_0(\mathbb{R})^2 $ with $ \psi_n \rightarrow \psi$ and
$H \psi_n \rightarrow H \psi$ as $n \rightarrow \infty$. This
shows that $\{ (H - \omega) \psi_n \}_{n \in \mathbb{N}} \subseteq
\algcal{D}(S_\omega)$ converges to $(H - \omega) \psi = \phi$. We conclude that
$\algcal{D}(S_\omega)$ is dense. We now calculate the operator product
$S_\omega \left(H-\omega
\right)$ on $C^\infty_0(\mathbb{R})^2$. For an arbitrary $\Psi =(\psi_1 ,
\psi_2)^T \in C^\infty_0(\mathbb{R})^2$ we
have
\begin{eqnarray*}
\left(S_\omega (H-\omega) \psi \right) (u) = \int_{-\infty}^{\infty}
k_\omega(u,v) (H-\omega) \psi (v) \: dv = \hspace{2 cm}\\
= \left(%
\begin{array}{c}
  0 \\
  -\omega \psi_1  + \psi_2 \\
\end{array}%
\right)(u) + \hspace{5.5cm}\\
+ \int_{-\infty}^\infty s_\omega (u,v) \left(%
\begin{array}{cc}
  - \partial_v ^2 + V_\omega(v) & 0 \\
  \omega\left( - \partial_v ^2 + V_\omega(v) \right)& 0 \\
\end{array}%
\right) \left(%
\begin{array}{c}
  \psi_1  \\
  \psi_2 \\
\end{array}%
\right)(v) \: dv \; .
\end{eqnarray*}
Hence, according to Lemma \ref{lemma zu function s(u,v)},
$$ S_\omega (H- \omega) = Id \quad \textrm{on } C^\infty_0
(\mathbb{R})^2 \; . $$ This yields that $S_\omega = (H- \omega)^{-1}$ on the
dense set $ \algcal{D}(S_\omega) $. Since $(H-\omega)^{-1}$ is a bounded
operator, the claim follows.
\end{proof}

As mentioned at the beginning of this section, we can now apply Stone's
formula for the spectral projections of $H$ and get for every $\Psi  \in
\algcal{H}$
\begin{eqnarray*}
\frac{1}{2} \left[ P_{[a,b]} + P_{(a,b)} \right] \Psi = \hspace{7 cm} \\
= \lim_{\epsilon \searrow 0} \frac{1}{2\pi i} \int_a^b \left[(H- \left(\omega +
i\epsilon)\right)^{-1} - \left(H- (\omega - i \epsilon)\right)^{-1} \right]\Psi
\: d \omega \; ,
\end{eqnarray*}
and this yields together with Proposition \ref{construction of the resolvent}
\begin{equation} \label{eq: stones formula in unserem fall}
= \lim_{\epsilon \searrow 0} \frac{1}{2\pi i} \int_a^b \left(
\int_{\mathbb{R}}
\left( k_{\omega + i \epsilon}(.,v) - k_{\omega - i \epsilon} (. ,v) \right)
\Psi (v) dv \right) d \omega \; ,
\end{equation}
where the limit is with respect to the norm in $\algcal{H}$.
It is therefore of special interest how the kernels $k_{\omega + i
\epsilon}(u,v)$ and $ k_{\omega -i \epsilon}(u,v)$ behave as
$\epsilon \searrow 0$. Since these kernels are given explicitly in terms of the
fundamental solutions $\acute{\phi}_{\omega \pm i \epsilon}$ and $
\grave{\phi}_{\omega \pm i \epsilon}$, we will discuss their behavior in the
next section.

\section{The Jost Solutions of the Radial Equation}

In this section we want to discuss the existence and the behavior of the
solutions $ \acute{\phi}_\omega , \grave{\phi}_\omega$ of the Schr\"odinger
equation (\ref{Schroedinger equation}), which in Section 4 we used for the
construction of the resolvent. We will prove the following theorem.

\begin{theorem} \label{Existenz der Loesungen acute phi grave phi}
\
\begin{enumerate}
    \item \label{Existenz von acute phi}
    For every $\omega \in  D = \left\{ \omega \in \mathbb{C} \: | \: \Im
    \omega \leq \frac{1}{4M} \right\}$, there exists a unique solution
$\phi_1(\omega,u)$ of the 
    Schr\"odinger equation (\ref{Schroedinger equation}) satisfying the boundary
conditions
    (\ref{boundarycond3}) such that for every fixed $u \in \mathbb{R}$ the function
    $\phi_1(\omega,u)$ is holomorphic in $\omega \in \overcirc{D}$ and
    continuous in $D$.

    \item  \label{Existenz von grave phi}
    For every angular momentum number $l$, the
    solutions $\grave{\phi}_\omega$ of the Schr\"odinger equation
(\ref{Schroedinger equation})
    with boundary conditions (\ref{boundarycond4})
    are well-defined and uniquely determined on the set $$ E= \left\{
    \omega \in \C \: \big| \: \Im \omega \leq 0, \: \omega \neq 0 \right\}
    \; . $$ For each fixed $u \in \R$, the function $ \grave{\phi}_\omega (u)$
is
    holomorphic in $\omega \in \overcirc{E}$ and continuous in $E$.

    Furthermore, in the case $l= 0$, $\grave{\phi}_\omega (u)$ may be
continuously
    extended to $\omega = 0$.
\end{enumerate}
\end{theorem}
Once having proven this theorem, we simply set
\begin{equation} \label{Defintion acute phi obere Halbebene}
\acute{\phi}_\omega (u) := \left\{ \begin{array}{cl}
    \overline{ \phi_1(\overline{\omega}, u)} & \textrm{, if } \Im \omega > 0 \\
    \phi_1(\omega,u) & \textrm{, if } \Im \omega \leq 0 \\
\end{array} \right. \; ,
\end{equation}
as well as
\begin{equation} \label{Definition grave phi obere Halbebene}
\grave{\phi}_\omega(u) := \overline{\grave{\phi}_{\overline{\omega}}(u)} \quad
\textrm{if } \Im \omega > 0\; ,
\end{equation} to obtain the solutions of Section 4. For $\Im \omega < 0$ this
is clear by definition. But in the case of $\Im \omega > 0$ the above defined
$\acute{\phi}_\omega(u), \grave{\phi}_\omega(u)$
are indeed the unique solutions of the Schr\"odinger equation
(\ref{Schroedinger equation}) with the desired boundary conditions
(\ref{boundarycond1}) and (\ref{boundarycond2}), respectively. This follows
immediately by complex conjugation of the Schr\"odinger equation due to the fact
that our potential $V_l$ is real.

\bigskip

For the proof of Theorem \ref{Existenz der Loesungen acute phi grave phi} we
will formally manipulate the Schr\"odinger equation with boundary conditions
(\ref{boundarycond3}, \ref{boundarycond4}) in order to get an appropriate
integral equation (which in different contexts is called the Jost or
Lipman-Schwinger equation). Then we will perform a perturbation expansion and
get estimates for all the terms of the expansion. A reference for this method
can be found e.g. in \cite[Section XI.8]{RS3}. Since this reference
contains only an outline of the proof, it seems worth working out the
details.

\bigskip

To introduce the method, we begin with the solutions $\phi_1(\omega,u)$. First
we write the Schr\"odinger equation (\ref{Schroedinger equation}) in the form
\begin{equation} \label{Herleitung Jost equation Schroedinger equation}
\left( - \frac{d^2}{du^2} - \omega^2 \right) \phi_\omega(u) = - W(u)
\phi_\omega(u) \; ,
\end{equation}
where $W$ is a potential in $L^1(\R)$ (later on, $W$ will be
replaced by $V_l$). Next we define for $\omega \in \C$ the function $
G_\omega (u)$ by
\begin{equation} \label{Greens funktion G}
G_\omega(u) := \left\{ \begin{array}{cl}
  0 & \textrm{, if } u \leq 0 \\
\displaystyle  - \frac{1}{\omega} \sin (\omega u) & \textrm{, if } u > 0
\textrm{ and } \omega \neq 0  \\
  - u & \textrm{, if } u > 0 \textrm{ and } \omega = 0 \quad .
\end{array} \right.
\end{equation}
A simple computation shows that $G_\omega (u)$ defines a \emph{Green's
function} for the operator on the left hand side of the equation
(\ref{Herleitung Jost equation Schroedinger equation}) in the sense that the
distributional equation
$$ \left(-\frac{d^2}{du^2} -\omega^2 \right) G_\omega (u) = \delta (u) $$
holds. In order to build in the boundary condition (\ref{boundarycond3}), we
make
in equation (\ref{Herleitung Jost equation Schroedinger equation}) the
substitution $\phi_\omega (u) = e^{i\omega u} + \tilde{\phi}_\omega(u) $ to
obtain
\begin{equation}
\left( - \frac{d^2}{du^2} - \omega^2 \right) \tilde{\phi}_\omega(u) = - W(u)
\phi_\omega(u) \; . \nonumber
\end{equation}
Solving this equation formally by convoluting the right hand side with
$G_\omega$, we get the formal solution
$$ \tilde{\phi}_\omega(u) = \big(( - W \phi_\omega) \ast G_\omega \big) (u)
\equiv
- \int_{-\infty}^\infty G_\omega (u-v) W(v) \phi_\omega(v) \: dv \; .$$ 
Hence $\phi_\omega(u) $ satisfies the equation 
\begin{equation} \label{Jost equation bound cond -unendlich}
\phi_\omega (u) = e^{i \omega u} - \int _{-\infty}^u G_\omega(u-v) W(v)
\phi_\omega(v) \:dv \; ,
\end{equation}
which is referred to as the \emph{Jost equation with boundary conditions at
$-\infty$}. Its significance lies in the fact that we can now easily perform a
perturbation expansion in the potential $W$. Namely, making for $\phi_\omega$
the ansatz as the perturbation series
\begin{equation} \label{Reihenentwicklung von phi 1}
\phi_\omega = \sum_{k=0}^\infty \phi_\omega^{(k)} \; ,
\end{equation}
we are led to the iteration scheme
\begin{equation} \label{Iterationsschema fuer phi 1}
\left. \begin{array}{rcl}
  \phi_\omega^{(0)} (u) & = & e^{i \omega u} \\
  \ & \vdots & \ \\
    \phi_\omega^{(k+1)}(u) & = & \displaystyle - \int_{-\infty}^u G_\omega(u-v)
W(v) \phi_\omega^{(k)}(v) \:dv \\
\end{array} \right\}  \begin{array}{c}
   \\
   \\
  . \\
\end{array}
\end{equation}
This iteration scheme can be used to construct solutions of the Jost equation.

We remark that under certain assumptions on $W$ like continuity,
the Jost equation is equivalent to the corresponding Schr\"odinger equation with
appropriate boundary conditions. We will show this for our special case $W
\equiv V_l$. A systematic method to rewrite second-order differential equations
with boundary conditions as integral equations can be found e.g. in
\cite[Section XI.8 Appendix 2]{RS3}.

We now state a theorem about solutions of the Jost equation. We consider
more general potentials $W$ than we have in our case, because it might be of
interest by itself.

\begin{theorem} \label{theorem ueber Loesungen der Jost equation}
Suppose that $W$ is a measurable function obeying for a given $u_0 < 0$ the 
condition $ \int_{-\infty}^{u_0}
|W(v)|dv < \infty$. Define for $u \leq u_0$ the function $P_\omega(u)$ by
\begin{equation} \label{Definition P_omega (u)}
P_\omega(u) = \int_{-\infty}^u \frac{4|v|}{1 + |\omega v|} \: |W(v)| e^{-
\left( \Im \omega + | \Im \omega | \right) v} \:dv \; .
\end{equation}
Then:
\begin{enumerate}
    \item
    For each $\omega \in E = \{ \omega \in \C \: \big| \: \Im \omega
     \leq 0 , \omega \neq 0 \}$ the Jost equation
     (\ref{Jost equation bound cond -unendlich}) has a unique solution
     $\phi_\omega (u)$ obeying $ \displaystyle \lim_{u \rightarrow -\infty}
\big| e^{-i \omega u}
     \phi_\omega (u) \big| < \infty $. Moreover, $\phi_\omega (u)$ is
continuously differentiable in $u$ on
     $ ( -\infty , u_0 )$ with $\displaystyle \lim_{ u \rightarrow -\infty} e^{
-i
     \omega u} \: \phi_\omega (u) = 1 $ and $\displaystyle \lim_{ u \rightarrow
-\infty} e^{ -i
     \omega u} \: \phi'_\omega (u) = i \omega$. For each fixed $u$, 
     the functions $\phi_\omega(u)$ and $\phi'_\omega (u)$
     are holomorphic in $\overcirc{E}$ and continuous in $E$. They satisfy the
     bounds
     \begin{eqnarray}
     \big| \phi_\omega (u) -  e^{i \omega u} \big| & \leq & e^{- u \,
     \Im \omega } \big| e^{P_\omega (u)} -1 \big| \label{Abschaetzung fuer
phi_omega(u)}
     \\
     \big| \phi'_\omega (u) - i \omega e^{i \omega u} \big| & \leq & e^{- u \,
     \Im\omega }  e^{P_\omega (u)}  \int_{- \infty}^u |W(v)| \: dv \; .
     \label{Abschaetzung fuer phi'_omega(u)}
     \end{eqnarray}

    \item
     If $\int_{-\infty}^{u_0} |v| |W(v)| \: dv < \infty $, for
every $u \leq u_0$ the function $\phi_\omega (u) $ may be continuously extended
to
$\omega = 0$. 
     Moreover, (\ref{Abschaetzung fuer     phi_omega(u)}),
     (\ref{Abschaetzung fuer phi'_omega(u)}) hold also at $\omega = 0$.

    \item
    If $\int_{-\infty}^{u_0} e^{- m v} |W(v)| dv < \infty$, for
every $u \leq u_0$ the function $\phi_\omega(u)$ can be extended to a
holomorphic
function in $\left\{ \omega \: | \: \Im \omega< \frac{1}{2}m \right\}$,
continuous in
$\left\{ \omega \: | \: \Im \omega \leq \frac{1}{2}m \right\}$. Moreover, in
the interval $ 0 < \Im \omega < \frac{1}{2}m$ the inequalities
(\ref{Abschaetzung fuer
phi_omega(u)}),(\ref{Abschaetzung fuer phi'_omega(u)}) are replaced by
\begin{eqnarray}
\big| \phi_\omega (u) - e^{i\omega u} \big| & \leq & \frac{1}{|\omega|} e^{u \Im
\omega} e^{P_\omega (u)} \int_{- \infty}^u e^{-2 v \Im \omega} |W(v)|\, dv
\label{Abschaetzung fuer phi_omega(u) Im>0} \hspace{1cm}\\
\big| \phi_\omega '(u) - i \omega e^{i\omega u} \big| & \leq &
e^{u \Im \omega} e^{P_\omega (u)} \int_{- \infty}^u e^{-2 v \Im \omega} |W(v)|\,
dv \label{Abschaetzung fuer phi'_omega(u) Im>0} \; .
\end{eqnarray}
\end{enumerate}
In each case, $\phi$ obeys $ \overline{\phi_\omega(u)} \equiv
\overline{\phi(\omega,u)} = \phi(-\overline{\omega},u)$. 
\end{theorem} 
We call this solution $\phi$ the \textbf{Jost solution}. 
For the proof of this theorem we need a good estimate for the Green's function
$G_\omega $.
\begin{lemma} \label{lemma zur abschaetzung fuer sinus}
For all $u \in \C$,
\begin{equation} \label{eq: Ungleichung fuer sin}
|\sin u| \: \leq \: \frac{2 |u|}{1+ |u|}\, e^{| \Im u |} \; .
\end{equation}
In particular, if $\omega \neq 0$ and $v \leq u \leq 0$,
\begin{equation} \label{eq: Schranke fuer G_omega}
\bigg| \frac{1}{\omega} \sin(\omega(u-v)) \bigg| \: \leq \: \frac{4|v|}{1 +
|\omega
v|} \: e^{-v |\Im \omega| - u \, \Im \omega} \; .
\end{equation}
\end{lemma}

\

\begin{proof}
In the case $|u| \geq 1$, the inequality (\ref{eq: Ungleichung fuer sin})
follows directly from the Euler formula $ \sin u = \frac{1}{2 i} \left(e^{i
u} - e^{-iu} \right)$ and the estimate
$$ (1 + |u|) |\sin u| \: \leq \: \frac{1}{2} (1 + |u|) 2 e^{|\Im u|} \: \leq \: 
2|u| e^{|\Im
u|} \; .$$

In the remaining case $|u| < 1$, we again use the Euler formula to obtain
$$ (1 + |u|) |\sin u| = \frac{1}{2} (1 + |u|) \big| e^{iu} - e^{-iu} \big| =
\frac{1}{2} (1 +|u|) \Big| \int_{-1}^1 iu e^{iu \tau} \: d\tau \Big| \; ,$$ and
hence
$$(1 + |u|) |\sin u| \: \leq \: \frac{1}{2} (|u| + |u|^2) \int_{-1}^1 \big|
e^{iu \tau} \big| \: d \tau
\: \leq \: \frac{1}{2} ( |u| + |u|^2) 2 \, e^{| \Im u|} \; .$$ Now (\ref{eq:
Ungleichung fuer sin}) follows by the assumption $|u| <1$.

In order to show (\ref{eq: Schranke fuer G_omega}) we use the identity
$$ \frac{1}{\omega} \sin(\omega(u-v)) =  \frac{1}{\omega} \left(
\sin (\omega u) e^{ i \omega v} - \sin (\omega v) e^{i \omega u} \right)$$ and
apply (\ref{eq: Ungleichung fuer sin}),
\begin{eqnarray}
 \bigg| \frac{1}{\omega} \sin(\omega(u-v))
\bigg| \: \leq \: \frac{1}{| \omega |}  \left( \big|\sin (\omega u) e^{ i \omega
v}
\big| +
\big|\sin (\omega v) e^{i \omega u} \big| \right) \nonumber\\
\leq^{^{^{\hspace{-4.7 mm} (\ref{eq: Ungleichung fuer sin})}}}  \frac{2 |u|}{1+
|\omega u|} \,  e^{|u \Im \omega|} e^{-v \Im \omega}  + \frac{2 |v|}{1+ |\omega
v|} \, e^{|v \Im \omega|} e^{-u \Im \omega} \; . \label{eq: Zwischenschritt im
Lemma zur abschaetzung von G}
\end{eqnarray}
Due to the assumption $0 \geq u \geq v$, we know that $|v| \geq |u|$ and thus
$$ \frac{2 |u|}{1+ |\omega u|} \leq \frac{2 |v|}{1+ |\omega v|}\; , \quad
 \; - u | \Im \omega | - v \Im \omega \leq  - v | \Im \omega | - u \Im \omega \;
. $$
Using these inequalities in (\ref{eq: Zwischenschritt im
Lemma zur abschaetzung von G}) the claim follows.
\end{proof}

Note that the estimate (\ref{eq: Schranke fuer G_omega}) remains valid in the
limit $0 \neq \omega \rightarrow 0$, if one replaces $\frac{1}{\omega} \sin
(\omega (u-v))$ by the function $u-v$.

Now we are ready to prove Theorem \ref{theorem ueber Loesungen der
Jost equation}:

\begin{proof} [Proof of Theorem \ref{theorem ueber Loesungen der Jost equation}]
\ \\
Using the perturbation expansion (\ref{Reihenentwicklung von phi 1}) together
with the iteration scheme (\ref{Iterationsschema fuer phi 1}), one easily sees
that we have already found a \emph{formal} solution. So our goal is to show
that this series is well-defined and has the desired properties. To this end,
we shall prove inductively that
\begin{equation} \label{eq: induktive Abschaetzung fuer iterationsschema}
\big| \phi^{(k)}_\omega (u) \big| \leq e^{- u \, \Im \omega} \frac{1}{k!}
P_\omega(u) ^k \quad \textrm{for all } k \in \N_0 ,\; \textrm{for all }\omega,u
\end{equation}
such that $P_\omega (u)$ is well-defined by (\ref{Definition P_omega (u)}).
Due to the integrability conditions on the potential $W$ in the statement of
the theorem this is the case for $u \leq u_0$ and for all $\omega \in E$ (cf.
$(i)$),
$\omega \in \overline{E}$ (cf. $(ii) $), $\omega \in \{\Im \omega \leq
\frac{1}{2}m\}$ (cf. $(iii)$), respectively. Furthermore, $P_\omega (u)$ is
continuous in
$u$ as well as in $\omega$ in these domains. The first statement is obvious
while the latter is due to the fact that the integrand in the definition
(\ref{Definition P_omega (u)}) is continuous in $\omega$ and one directly finds
an integrable dominating function such that one can apply Lebegue's Dominated
Convergence Theorem.

\bigskip

We start the induction with the case $k=0$ for which (\ref{eq: induktive
Abschaetzung fuer iterationsschema}) certainly is satisfied.
Thus assume that (\ref{eq: induktive Abschaetzung fuer iterationsschema}) holds
for a given $k$. Then, estimating the integral equation in
(\ref{Iterationsschema fuer phi 1}) using (\ref{eq: Schranke fuer G_omega}) and
(\ref{Definition P_omega (u)}), we
obtain
\begin{eqnarray*}
\big| \phi_\omega^{(k+1)}(u) \big|  &  \leq  &  \int_{-\infty}^u
|G_\omega(u-v)| |W(v)| \big|\phi_\omega^{(k)}(v) \big| \:dv \\
 & \leq &
 \int_{-\infty}^u \frac{4 |v|}{ 1 + | \omega v|} \: e^{-v |\Im \omega| - u \,
\Im
 \omega} |W(v)| e^{-v \Im \omega}  \frac{1}{k!} \, P_\omega (v)^k \: dv \\
 & = & e^{-u \Im \omega} \, \frac{1}{k!}
\int_{-\infty}^u
 \frac{ d P_\omega}{dv} (v)  \, P_\omega (v)^k \: dv \\
 & = & e^{-u \Im \omega} \frac{1}{(k+1)!} P_\omega (u) ^{k+1} \; ,
\end{eqnarray*}
where in the last step we used that $P_\omega(u)$ vanishes when $u$ goes to
$-\infty$. This concludes the proof of (\ref{eq: induktive
Abschaetzung fuer iterationsschema}).

Summing over $k$, (\ref{eq: induktive Abschaetzung fuer iterationsschema})
yields the inequality
\begin{equation} \label{eq: Summation der Abschaetzungen}
\sum_{k=0}^{\infty} \big| \phi_\omega^{(k)} (u) \big| \leq e^{-u \Im \omega}
e^{P_\omega (u)}.
\end{equation}
Because of the continuity of $P_\omega(u)$, the series (\ref{Reihenentwicklung
von phi 1}) converges uniformly for $u$ and $\omega$ in compact sets. Using the
iteration scheme (\ref{Iterationsschema fuer phi 1}), this series can
be written as
$$ \sum_{k=0}^\infty \phi_\omega^{(k)} (u) = e^{i \omega u} - \sum_{k=1}^\infty
 \int_{-\infty}^u G_\omega (u-v) W(v) \phi_\omega^{(k-1)} (v) \, dv \; ,$$ and
the bound (\ref{eq: Summation der Abschaetzungen}) allows us to apply Lebesgue's
dominated convergence theorem and to interchange the sum and the integral.
Hence the series is indeed a solution of the Jost equation (\ref{Jost equation
bound cond -unendlich}).

\bigskip

Next we want to show that a solution of the Jost equation is continuously
differentiable with respect to $u$. To this end, we first compute for an
arbitrary $u < u_0$ the difference quotient,
\begin{eqnarray}
\frac{1}{h} \left(\phi_\omega(u + h) - \phi_\omega (u) - e^{i \omega
(u+h)} + e^{i\omega u} \right) \overeq{(\ref{Jost
equation bound cond -unendlich})} \hspace{2.7 cm} \nonumber \\
\int_{- \infty}^{u+h} \frac{1}{h \omega} \left[ \sin(\omega (u +h -v)) - \sin(
\omega (u-v)) \right] W(v) \phi_\omega (v) \, dv +
\label{eq:DifferenzenquotientIntegral 1} \\
+ \frac{1}{h} \int_u^{u+h} \frac{1}{\omega} \sin(\omega (u-v)) W(v) \phi_\omega
(v) \, dv \; , \label{eq: Differenzenquotient Integral 2} \hspace{3.3
cm}
\end{eqnarray}
where $ h \neq 0 $. We may restrict attention to the case $\Im \omega
\leq 0$ and $h>0$ (the other cases are analogous). Using the
estimate
\begin{eqnarray*}
 \Big| \partial_u \left( \frac{1}{\omega} \sin
(\omega (u-v))\right) \Big| = \big| \cos (\omega (u-v)) \big| \leq \hspace{30mm}
\\
\leq \frac{1}{2} \left ( e^{-u \Im \omega + v \Im \omega} + e^{u\Im \omega - v
\Im \omega} \right)
\end{eqnarray*}
together with (\ref{eq: Summation der Abschaetzungen}), we can apply the mean
value theorem to the first integrand to obtain the dominating
function
$$ \frac{1}{2}\left( e^{-\xi(v) \Im \omega} e^{v \Im \omega} + e^{\xi(v) \Im
\omega}
 e^{ - v \Im \omega} \right) | W(v) | \: e^{-v \Im \omega} e^{P_\omega(v)} \;
 ,$$
where $\xi(v) \in [u,u+h]$. Due to the integrability conditions on $W$, it is
clear that this function is integrable. Hence Lebesgue's dominated convergence
theorem allows us to take
the limit $h\rightarrow 0$ in (\ref{eq:DifferenzenquotientIntegral 1}). This
gives
\begin{equation}
\nonumber \int_{- \infty}^u \cos (\omega(u-v)) W(v) \phi_\omega (v) \,dv \; .
\end{equation}
In order to treat the second integral, we choose $h<h_0$,
where $h_0$ is so small that
$$ \max_{v \in [u,u+h]} \Big| \frac{1}{\omega h} \sin (\omega(u-v)) \Big| \leq 
2  \quad \textrm{for all } h < h_0. $$ 
This is possible because 
$\displaystyle \lim_{h \rightarrow 0} \frac{1}{\omega h} \sin (\omega h) = 1$.
Thus we can estimate (\ref{eq: Differenzenquotient Integral 2}) by
\begin{eqnarray*}
\Big| \frac{1}{h} \int_u^{u+h} \frac{1}{\omega} \sin(\omega (u-v)) W(v)
\phi_\omega (v) \, dv \Big| \leq \hspace{2.5cm}\\
\leq 2 \, e^{-(u+h) \Im \omega}
e^{P_\omega (u+h)} \int_{(-\infty,u_0)} \big| 1_{[u,u+h]}(v) W(v) \big|\, dv
\; ,
\end{eqnarray*}
and the last integral goes to $0$ as $h \rightarrow 0$ by Lebesgue's monotone
convergence theorem using the
fact that $W \in L^1(-\infty, u_0)$. Hence (\ref{eq: Differenzenquotient
Integral 2}) vanishes.

Alltogether we conclude that $\phi_\omega(u)$ is differentiable with derivative 
\begin{equation} \label{eq: Ableitung von phi omega}
\phi_\omega'(u) = i \omega e^{i \omega u} + \int_{- \infty}^u \cos (\omega
(u-v)) W(v) \phi_\omega(v)dv \; ,
\end{equation}
which is continuous on $(-\infty, u_0)$ because of the estimate (\ref{eq:
Summation der Abschaetzungen}).

\bigskip

The estimate (\ref{Abschaetzung fuer phi_omega(u)}) is a simple consequence of
(\ref{eq: induktive Abschaetzung fuer iterationsschema}) together with the
perturbation expansion (\ref{Reihenentwicklung von phi 1}).
For the proof of (\ref{Abschaetzung fuer phi'_omega(u)}) we use the
representation of the
derivative (\ref{eq: Ableitung von phi omega}) together with the inequality
(\ref{eq: Summation der Abschaetzungen}):
\begin{eqnarray*}
\big| \phi_\omega ' (u) - i \omega e^{i \omega u} \big| \; \;
\overleq{(\ref{eq: Ableitung von phi omega})}  \int_{- \infty}^u |\cos (\omega
(u-v))| \, |W(v)|
\, |\phi_\omega (v)| dv \hspace{2cm} \\
 \overleq{(\ref{eq: Summation der Abschaetzungen})}   \int_{-\infty}^u
\frac{1}{2} \left( e^{- u \Im \omega} e^{v \Im \omega} + e^{ u \Im \omega} e^{-
v \Im \omega} \right) | W(v) | \, e^{-v \Im \omega} e^{P_\omega (v)} \, dv \\
\leq \; \; e^{-u \Im \omega} e^{P_\omega(u)} \int_{-\infty}^u |W(v)| \, dv \; ,
\hspace{5.1cm}
\end{eqnarray*}
where in the last step we used the fact that $P_\omega (v)$ and $e^{-v
\Im \omega}$ (with $\Im \omega \leq 0$) are monotone increasing. The estimates
(\ref{Abschaetzung fuer phi_omega(u) Im>0}) as well as
(\ref{Abschaetzung fuer phi'_omega(u) Im>0}) are shown in the same way.

\bigskip

Let us now verify that for any fixed $u$, the function $\phi_\omega(u)$ is
holomorphic in $\omega$, and
continuous on the domains as specified in $(i),(ii)$ and
$(iii)$. Due to the locally uniform convergence of the perturbation series, it
suffices to show that every $\phi_\omega^{(k)}(u)$ has the desired properties.
We do this inductively, where the case $k=0$ is
trivial. Let us now assume that $\phi_\omega^{(k)}(u)$ is holomorphic in
$\overcirc{E}$
($\{  \Im \omega < \frac{1}{2} m \}$, respectively). In order to prove
that $\phi_\omega^{(k+1)}$ is holomorphic, we want to apply Morera's theorem.
Thus we must show that $\phi_\omega^{(k+1)}(u)$ is continuous in
$\omega$ and that the integral
\begin{equation} \label{eq: Holomorphiebedingung bei Morera} 
\oint_{\gamma} \phi_\omega^{(k+1)} (u) \, d \omega
\overeq{(\ref{Iterationsschema fuer phi 1})} \oint_\gamma \int_{-\infty}^u
\frac{1}{\omega} \sin (\omega (u-v)) W(v) \phi_\omega^{(k)}(v) \, dv \, d\omega
\end{equation} 
vanishes for every closed contour $\gamma$ in $\overcirc{E}$ (or in case
$(iii)$, for every contour in $\{ \Im \omega < \frac{1}{2} m \}$, respectively).
Using the above estimates (\ref{eq: Schranke fuer G_omega}),(\ref{eq: induktive
Abschaetzung fuer iterationsschema}) together with the monotonicity of
$P_\omega (u)$ in $u$ we get the following bound for the
integrand
\begin{eqnarray}
 \Big| \frac{1}{\omega} \sin (\omega (u-v)) W(v) \phi_\omega^{(k)}(v) \Big| \leq
\hspace{4cm} \nonumber \\
\leq |W(v)| \frac{4|v|}{1 + |\omega v|} e^{-u \Im \omega -v |\Im \omega| -v \Im
\omega} \frac{1}{k!} P_\omega(u)^k \; . 
\label{eq: abschaetzung des integranden beim holomorphiebeweis}
\end{eqnarray}
Due to the induction
hypothesis, the integrand is continuous in $\omega$. Moreover, for a compact
neighborhood $K(\omega_0)$ of a fixed $\omega_0$ contained in the specified
domains, (\ref{eq: abschaetzung des integranden beim holomorphiebeweis}) yields
for the family $\frac{1}{\omega} \sin(\omega (u-v)) W(v)
\phi_\omega^{(k)} (v), \; \omega \in K(\omega_0)$ the uniformly dominating
function
$$  |W(v)| \frac{4|v|}{1 +| v| \min|\omega|\: } e^{-u
\Im \omega - v \: \max ( |\Im \omega| + \Im
\omega)} \frac{1}{k!} P_\omega(u)^k\; , $$
where the minimum and the maximum are taken in $K(\omega_0)$. This function is
integrable for $K(\omega_0)$ chosen sufficiently small due to the integrability
conditions on $W$. This lets us apply Lebesgue's dominated convergence theorem
to show the
continuity in $\omega$ for $\phi_\omega^{(k+1)}(u)$, which is given by the
integral
(\ref{Iterationsschema fuer phi 1}).
Moreover, (\ref{eq: abschaetzung des integranden beim holomorphiebeweis})
together with the continuity in $\omega$ of $P_\omega(u)$ yield that the
integral
\begin{equation*} 
 \oint_\gamma
\int_{- \infty}^u \Big| \frac{1}{\omega} \sin(\omega (u-v)) W(v)
\phi_\omega^{(k)}(v) \Big| \,dv \: d\omega < \infty
\end{equation*}
exists for an arbitrary closed contour $\gamma$ in $\overcirc{E}$ (or $\{ \Im
\omega < \frac{1}{2} m \}$, respectively). By the theorem of Fubini,
we may interchange the orders of integration in (\ref{eq: Holomorphiebedingung
bei Morera}). Because of the induction hypothesis, the integrand of
(\ref{eq: Holomorphiebedingung bei Morera}) on the right hand side is
holomorphic. Thus the integral vanishes due to the Cauchy integral
theorem. We conclude that $\phi_\omega^{(k)}$ is holomorphic for every $k$.
Since $\phi_\omega(u)$ is holomorphic, the same argument together with equation
(\ref{eq: Ableitung von phi omega}) yields that $\phi_\omega '$ is
also holomorphic.

\bigskip

It remains to prove uniqueness. Let $\psi_\omega (u)$ be another solution
of the Jost equation obeying $ \displaystyle \lim_{ u \rightarrow -\infty} \big|
e^{-i \omega u} \psi_\omega (u) \big|  <  \infty $. Then we can find a $c>0$
with $ \big| \psi_\omega (u) \big| \leq c e^{- u \Im \omega}$ for all $u \leq
u_0$. Then as above one shows inductively that
$$ \Big| \psi_\omega (u) - \sum_{l=0}^N
\phi_\omega^{(k)} (u) \Big| \leq c \, e^{- u \Im \omega} \frac{1}{(N+1)!}
P_\omega (u) ^{N+1} \; ,$$ and taking $N \rightarrow \infty$ we obtain
$\psi_\omega = \phi_\omega$. \\
The uniqueness also implies that $ \overline{\phi(\omega,u)} =
\phi(-\overline{\omega},u)$, concluding the proof.
\end{proof}

\begin{remark} \label{rem: analoges theorem zu jost equation}
In order to treat the Schr\"odinger equation (\ref{Schroedinger equation}) with
boundary conditions at infinity (\ref{boundarycond4}), we derive the
corresponding Jost equation with boundary equations at $+\infty$ using
the same procedure as on page
\pageref{Herleitung Jost equation Schroedinger equation}:
\begin{equation}
\label{eq: Jost gleichung randbedingungen +unendlich}
\phi_\omega (u) = e^{-i\omega u} - \int_u^{\infty} \frac{1}{\omega} \sin (\omega
(u
-v)) W(v) \phi_\omega (v) \, dv \; .
\end{equation}
It is obvious that the solution $\tilde{\phi}_\omega(u)$ of the Jost equation
with boundary \linebreak conditions at $-\infty$ with potential $W(-v)$
constructed in Theorem \ref{theorem ueber Loesungen der Jost equation} gives
rise to a solution $\phi_\omega$ of (\ref{eq: Jost gleichung randbedingungen
+unendlich}) by defining $\phi_\omega (u) := \tilde{\phi}_\omega (-u) $.
\end{remark}

\bigskip

With the results of Theorem \ref{theorem ueber Loesungen der Jost equation} it
is now easy to prove Theorem \ref{Existenz der Loesungen acute phi grave phi}:

\begin{proof} [Proof of Theorem \ref{Existenz der Loesungen acute phi grave
phi}]
Let us apply Theorem \ref{theorem ueber Loesungen der Jost equation} to the
potential $ V_l(u) $ given by (\ref{potential}), which is obviously a smooth
function in $u$. Furthermore, it vanishes on the event horizon $2M$ with the
asymptotics $V_l= \mathcal{O} (r-2M)$. Using the definition of the Regge-Wheeler
coordinate $u$ (\ref{reggewheelercoord}), this means that $V_l(u)$ decays
exponentially as $u \rightarrow -\infty$. More precisely, there is a constant
$c> 0$ such that 
\begin{equation*}
\displaystyle | V_l(u)| \leq c \, e^{\frac{u}{2M}} \quad \textrm{for small } u
\;.
\end{equation*}
Theorem \ref{theorem ueber Loesungen der Jost equation} $(iii)$ yields for
$u\leq u_0 <0 $ a solution $ \phi_1 (\omega,u)$ of the Jost equation (\ref{Jost
equation bound cond -unendlich}) with the desired properties. It
remains to show that $\phi_1$ is also a solution of
the Schr\"odinger equation (\ref{Schroedinger equation}) for $u\leq u_0$. (Due
to the Picard-Lindel\"of theorem, this solution of the linear equation can be
uniquely extended to $u \in \R$; the resulting function is analytic
in $\omega$ due to the analytical dependence in $\omega$ from the coefficients
and initial conditions.) But this follows immediately by differentiating
equation (\ref{eq: Ableitung von phi omega}) and using that $V_l \equiv W$ is
smooth, so that the whole integrand is at least differentiable with respect
to $v$. We have then proven the existence of $\acute{\phi}_\omega$. For the
uniqueness, we show that in our special case every solution of
(\ref{Schroedinger equation}) with boundary conditions (\ref{boundarycond3})
is a solution of (\ref{Jost equation bound cond -unendlich}). This can be done
by integration by parts: For let $\psi_\omega (u)$ be such a solution. Then
\begin{eqnarray*}
\int_{-\infty}^u\frac{1}{\omega} \sin(\omega (u-v)) V_l(v) \psi_\omega (v) \,
dv = \hspace{4.5cm} \\ 
= \int_{-\infty}^u \frac{1}{\omega}\sin(\omega(u-v))(\partial_v ^2 +\omega^2)
\psi_\omega (v) \, dv = \psi_\omega(u) - e^{i \omega
u} \; ,
\end{eqnarray*}
where the remaining terms vanish due to the boundary conditions. Since we know
that the solution of the Jost equation is uniquely determined, this must be also
the case for the solution of the Schr\"odinger equation. Thus we have proven
part
$(i)$.

For the proof of $(ii)$ we refer to Remark \ref{rem: analoges theorem zu jost
equation}. In contrast to the exponential decay at $- \infty$, the
potential $V_l(u)$ has only polynomial decay at $+ \infty$. More precisely,
according to the definition of $u$, $V_l(u) = \mathcal{O} (\frac{l(l+1)}{u^2})$
for $l \geq 1$, \linebreak$V_0 (u) = \mathcal{O} (\frac{2M}{u^3}) $,
respectively, as $u
\rightarrow \infty$. Thus we can apply the analogs of Theorem
\ref{theorem ueber Loesungen der Jost equation} $(i)$, $(ii)$, respectively.
This
gives the existence and uniqueness of the solution $\grave{\phi}_\omega$
for the Schr\"odinger equation with the stated properties.
\end{proof}

When taking the limit $\epsilon \searrow 0$ in Stone's formula (\ref{eq: stones
formula in unserem fall}), the behavior of $\grave{\phi}_\omega (u)$ at
$\omega = 0$ still causes problems. While in the case $l=0$ we know from Theorem
\ref{Existenz der Loesungen acute phi grave phi} that $\grave{\phi}_\omega$ 
can be continuously extended there, we do not yet know what happens for $l \neq
0$. The following theorem settles this problem by showing that, after suitable
rescaling, the solutions $\grave{\phi}_\omega$ have a well-defined
limit at $\omega = 0$:

\begin{theorem} \label{theorem: Konvergenz der Loesungen bei omega = 0}
For every angular momentum number $l$, there is a solution $\phi_0$ of the
Schr\"odinger equation (\ref{Schroedinger equation}) for $\omega = 0$ with the
asymptotics
\begin{equation} \label{eq: asymptotik der Loesung der SG fuer omega = 0}
\lim_{u \rightarrow \infty}  u^l \phi_0 (u) = i^l
\frac{2^l \sqrt{\pi} } {\Gamma(\frac{1}{2} - l)} = (-i)^l (2l-1)!! \; ,
\end{equation}
where $$(2l-1)!! := \left\{ \begin{array}{cll} (2l - 1) \cdot (2l-3) \cdot ...
\cdot 3 \cdot 1 & , & \textrm{if } l \neq 0 \\ 1  & , & \textrm{if } l= 0 \; .
\end{array}\right.$$
This solution can be obtained as a limit of the solutions from Theorem
\ref{Existenz der Loesungen acute phi grave phi}, in the sense that for all $u
\in \R $,
\begin{equation} \label{eq: Stetigkeit omegahochlphiomega gegen phi0}
\phi_0 (u) = \lim_{E \ni \omega \rightarrow 0} \omega^l
\grave{\phi}_\omega (u) \quad \textrm{and} \quad  \phi_0 '(u) = \lim_{E \ni
\omega \rightarrow 0} \omega^l \grave{\phi}'_\omega (u) \; .
\end{equation}
\end{theorem}
Note that the above properties of the solution $\phi_0$ really
coincide in the case $l = 0$ with that of the solution $\grave{\phi}_0$ already
constructed in
Theorem \ref{Existenz der Loesungen acute phi grave phi} $(ii)$.

\bigskip

For the proof of this theorem we use the same method as in the proof for
Theorem \ref{Existenz der Loesungen acute phi grave phi}. However, the iteration
scheme (\ref{Iterationsschema fuer phi 1}) does not work for $l \neq 0$ in the
limit $\omega \rightarrow 0$, because the integral 
$$ \phi_0^{(1)} (u) = - \int_u^{\infty} (u-v) V_l(v) \: dv  $$
diverges ($V_l(u)$ decays only quadratically at infinity for $l \neq 0$). We
avoid this problem by adding the leading asymptotic term of the potential $V_l$
to the unperturbed equation,
\begin{equation} \label{eq: stoerungsgleichung fuer schroedinger gleichung 2}
\left ( - \frac{d^2}{d u^2} - \omega^2 +\frac{(l+ \frac{1}{2})^2
-\frac{1}{4}}{u^2} \right) \phi_\omega (u) = - W_l(u) \phi_\omega (u)\; .
\end{equation}
Now the perturbation term $W_l(u) = V_l(u) -\frac{l(l+1)}{u^2}$
has the asymptotics $W_l(u) = \mathcal{O} (\frac{\log u}{u^3}) $.

Fortunately, the unperturbed differential equation
corresponding to (\ref{eq: stoerungsgleichung fuer schroedinger gleichung 2})
can still be solved exactly. The solutions can be expressed in terms of Bessel
functions.
For our further consideration, the two functions
\begin{equation} \label{eq: Loesungen der stoerungsfreien DGL mit
Besselfunktionen}
h_1(l,\omega, u) = \sqrt{\frac{\pi \omega u}{2}} J_{l+\frac{1}{2}} (\omega u)\;
, \quad h_2(l,\omega, u) = \sqrt{\frac{\pi \omega u}{2}} J_{-l-\frac{1}{2}}
(\omega u)
\end{equation}
play an important role. Here the function $J_\nu (u)$ is the Bessel
function of the first kind (a good reference for the theory of the Bessel
functions is \cite{Watson}). It solves Bessel's
differential equation 
\begin{equation}
\nonumber  u^2 y'' (u) + u y'(u) + (u^2 - \nu^2) y(u) = 0 \;.
\end{equation}
In addition, it is an analytic function in $\nu$ and $u$ for all values
of $\nu$ and $u \neq 0$ (if $\Re \nu \geq 0$, it can be analytically extended 
even to $u = 0$). It has the series expansion
\begin{equation} \label{eq: Reihendarstellung der Besselfunktion}
J_\nu (u) = \sum_{m= 0}^\infty \frac{(-1)^m}{m! \Gamma (\nu +m +1)} \left(
\frac{u}{2}\right)^{\nu +2m} 
\end{equation}
and the following asymptotics for $ |u| \gg 1$ (cf. \cite{Watson} 7.21):
\begin{eqnarray} \label{eq: Asymptotik der Besselfunktion}
J_\nu (u) \sim \sqrt{\frac{2}{\pi u}} \left[ \cos \left( u - \frac{\pi}{2} 
\left(
\nu +\frac{1}{2} \right) \right) \cdot \sum_{m=0}^\infty 
\frac{(-1)^m (\nu,2m)}{(2u)^{2m}} \nonumber \right.\\
\left. -\sin\left( u - \frac{\pi}{2} 
\left( \nu +\frac{1}{2} \right) \right) \cdot \sum_{m=0}^\infty 
\frac{(-1)^m (\nu,2m+1) }{(2u)^{2m+1}}  \right]  \; ,
\end{eqnarray}
where we have used the notation 
$$ (\nu,m) := \frac{\Gamma \left( \nu +m +\frac{1}{2} \right)}{m! \Gamma \left(
\nu - m +\frac{1}{2} \right)} \; .$$
Moreover, the derivatives satisfy the recurrence formulas
\begin{eqnarray*}
u J_\nu '(u) & = & u J_{\nu -1}(u) - \nu J_\nu (u) \quad \textrm{and} \\
u J_\nu '(u) & = & \nu J_\nu (u) - u J_{\nu +1} (u) \; .
\end{eqnarray*}
The Wronskian of the functions $J_\nu ,J_{- \nu}$ (which both
solve the same differential equation, since Bessel's differential equation is
symmetric in $\nu$) is given by the formula
\begin{equation} \label{eq: Wronskian von J nu J -nu}
w \left(J_\nu (u) ,J_{- \nu} (u) \right) = - \frac{2 \sin (\nu \pi)}{\pi u}\; .
\end{equation}
This yields that these functions form a fundamental system for Bessel's
differential equation provided that $\nu$ is not an integer.

\bigskip

In our applications we choose $\nu = l+ \frac{1}{2}$. Thus the functions $h_1(l,
\omega,u)$ and $h_2 (l, \omega, u)$ have the following asymptotics,
\begin{eqnarray} \label{eq: Asymptotik von h1}
h_1 (l,\omega,u) & \sim & \left\{
\begin{array}{lcl}
 \displaystyle \cos \left(\omega u - (l+1) \frac{\pi}{2} \right) & , &
\textrm{if }
|\omega u| \gg 1 \vspace{2 mm}\\ 
\displaystyle
\frac{\sqrt{\pi}}{\Gamma (\frac{3}{2} + l)} \left(\frac{\omega u}{2} \right)^{l+
1} & , & \textrm{if } |\omega u| \ll 1 
\end{array} \right \} \; \\ 
 \label{eq: Asymptotik von h2}
h_2 (l,\omega,u) & \sim & \left\{
\begin{array}{lcl}
 \displaystyle \cos \left(\omega u + l \frac{\pi}{2} \right) \hspace{4.4mm} & ,
& \textrm{if }
|\omega u| \gg 1 \vspace{2 mm}\\ 
\displaystyle
\frac{\sqrt{\pi}}{\Gamma (\frac{1}{2} - l)} \left(\frac{\omega u}{2}
\right)^{-l} \hspace{4.4mm} & , & \textrm{if } |\omega u| \ll 1 
\end{array} \right \} \; . \hspace{5 mm}
\end{eqnarray}
Furthermore, the formula (\ref{eq: Wronskian von J nu J -nu}) for the Wronskian
simplifies to
\begin{equation} \label{eq: Wronski von h1 und h2}
w\big(h_1(l,\omega,u), h_2(l,\omega,u)\big) = (-1)^{l+1}  \omega \; , \qquad
\textrm{if
}l \textrm{ is an integer },
\end{equation}
and this yields that in the
case $\omega \neq 0$ the solutions $h_1,h_2$ form a fundamental system.

Thus for $\omega \neq 0$ we take as the Green's function for the operator on the
left hand side of (\ref{eq: stoerungsgleichung fuer schroedinger gleichung 2})
the standard formula 
\begin{equation} \label{eq: greens funktion fuer stoerungsgleichung 2}
S_\omega (u,v) = \Theta(v- u) \frac{1}{w(h_1,h_2)} \big(h_1 (v)
h_2(u) - h_1 (u) h_2(v) \big) \; ,
\end{equation}
where $h_{1/2}(u) \equiv h_{1/2}(l,\omega, u)$ and $\Theta $ denotes
the Heaviside function defined by $\Theta (x) =1 $ if
$x \geq 0$ and $\Theta (x) = 0$ otherwise. Note that $S_\omega$ is also
well-defined in the limit $\omega \rightarrow 0$.
For this we use the asymptotics and the value of the Wronskian and get for very
small $\omega$,
\begin{eqnarray*}
\lim_{\omega \rightarrow 0} S_\omega (u,v) = \lim_{\omega \rightarrow 0}
\frac{(-1)^{l+1}}{\omega} \cdot \frac{\pi \omega}{2 \Gamma \left(\frac{3}{2} +
l\right) \Gamma \left(\frac{1}{2} -l \right)} \left( v^{l+1} u^{-l} - u^{
l+1} v^{-l} \right)  \\
= \frac{(-1)^{l+1}\pi }{2 \left(\frac{1}{2} + l\right) \Gamma \left(\frac{1}{2}
+ l\right) \Gamma \left(\frac{1}{2} -l \right)} \left( v^{l+1} u^{-l} - u^{
l+1} v^{-l} \right) = \hspace{2.3 cm}  \\ 
= \frac{(-1)^{l+1}\pi \cos (\pi l) }{(2l +1) \pi } \left( v^{l+1} u^{-l} -
u^{l+1} v^{-l} \right) = \hspace{4 cm} \\ 
= - \frac{1}{2l+1} \left( v^{l+1} u^{-l} - u^{l+1} v^{-l} \right) \; 
, \hspace{5.7 cm}
\end{eqnarray*}
where we have used some elementary properties of the Gamma function. This
also shows that the Green's function converges to the Green's function
$S_0(u,v)$ given by the above formula for the solutions $u^{l+1} , u^{-l}$ of
the unperturbed differential operator on the left hand side of (\ref{eq:
stoerungsgleichung fuer schroedinger gleichung 2}) for $\omega = 0$. \\
We now proceed with the perturbation series ansatz
\begin{equation} \label{eq: reihenansatz fuer integralgleichung im Besselfall}
\phi_\omega (u) = \sum_{m= 0}^\infty \phi_\omega^{(m)} (u) \; ,
\end{equation}
which, as at the beginning of this section, leads to the iteration scheme
\begin{equation} \label{eq: Iterationsschema fuer integralgleichung im
Besselfall}
\phi_\omega^{(m+1)} (u) = - \int_u^\infty S_\omega (u,v) W_l(v)
\phi_\omega^{(m)} (v) \:dv \; .
\end{equation}
As initial function we take 
$$ \phi_\omega^{(0)}(u) = \omega^l e^{-i (l+1)\frac{\pi}{2}} \sqrt{\frac{\pi
\omega u}{2}} H_{l+ \frac{1}{2}}^{(2)} (\omega u) \; ,$$
where $H_\nu^{(2)}$ is another solution of Bessels equation (called Bessel
function of the third kind or second Hankel function). It is related to $J_\nu$
by
$$  H_\nu^{(2)} (u) = \frac{J_{- \nu}(u) - e^{\nu \pi i } J_\nu (u) }{- i
\sin(\nu \pi)} \; ,$$ and has for large $|u|$ the asymptotics 
\begin{equation} \label{eq: Asymptotik der zweiten Hankelfunktion}
H_\nu ^{(2)} (u) \sim \sqrt{\frac{2}{\pi u}} e^{-i\left( u- \frac{1}{2}  \pi
\left( \nu + \frac{1}{2} \right) \right)} \sum_{m=0}^\infty
\frac{(\nu,m)}{(2iu)^m} \; .
\end{equation} 
Thus our intial function $\phi_\omega^{(0)} (u)$ solves the
unperturbed equation, and we have the relation 
\begin{equation} \label{eq: anfangsfunktion fuer Integralgleichung im
Besselfall}
\phi_\omega^{(0)} (u) = \omega ^l \left( (-i)^{l+1} h_1(l,\omega,u) + i^l h_2
(l,\omega,u) \right) 
\end{equation}
together with the asymptotics
\begin{equation} \label{eq: Asymptotik der Anfangsfunktion}
\phi_\omega^{(0)} (u) = \omega^l e^{-i \omega u} \left(1 + \mathcal{O} \left(
\frac{1}{u} \right) \right) \quad \textrm{, if } |u| \gg 1\; .
\end{equation}
Moreover, the function $ \phi_\omega^{(0)}$ converges in the limit $ \omega
\rightarrow 0$ pointwise for all $u \geq
u_0 >0 $:
\begin{equation} \label{eq: anfangsfunktion im limes omega gegen 0}
\lim_{\omega \rightarrow 0} \phi_\omega^{(0)} (u) = i^l
\frac{\sqrt{\pi}}{\Gamma \left( \frac{1}{2} -l \right) } \left( \frac{u}{2}
\right)^{-l} \; .
\end{equation}
Since we are interested in statements for $\omega = 0$, it is
convenient in what follows to restrict $\omega$ to the domain
$$ F := \{ \omega \in \C \: \big| \: \Im \omega \leq 0 , | \omega | \leq 1 \}
\;.$$
The following lemma yields that our perturbation series (\ref{eq: reihenansatz
fuer integralgleichung im Besselfall}) is well-defined.

\begin{lemma} \label{lemma: Konvergenz des Iterationsschemas}
For every $ u_0 >0 $, the iteration scheme
(\ref{eq: reihenansatz fuer integralgleichung im
Besselfall}),(\ref{eq: Iterationsschema fuer integralgleichung im
Besselfall}),(\ref{eq: anfangsfunktion fuer Integralgleichung im Besselfall})
converges locally uniformly for all $u\geq u_0$ and $\omega \in F$. In
particular, the functions $\phi_\omega (u)$ are for fixed $u$ a continuous
family in $\omega \in F$. They satisfy the integral equation
\begin{equation} \label{eq: Integralgleichung fuer phi omega im besselfall}
\phi_\omega (u) = \phi_\omega^{(0)} (u) - \int_u^\infty S_\omega(u,v) W_l (v)
\phi_\omega (v) \:dv \; .
\end{equation}
\end{lemma}

\begin{proof}
In order to prove the lemma, we need to derive good bounds for the initial
function $\phi_\omega^{(0)} (u)$ as well as for the Green's function $S_\omega
(u,v)$. To this end, we exploit the asymptotics of $h_1,h_2$. We thus obtain
the bound
\begin{equation} \label{eq: Abschaetzung fuer anfangsfunktion}
\frac{1}{C_1} \leq | \phi_\omega^{(0)} (u) | e^{-u \Im \omega } \left(
\frac{u}{1+ | \omega| u} \right)^l \leq C_1 \; .
\end{equation}
Likewise, for the Green's function we have (note
that $v\geq u >0$),
$$\begin{array}{lrcll} 
  & | S_\omega (u,v)|& \leq & \displaystyle C_2 \left(\frac{u}{1 + |\omega|
u} \right)^{-l}
\left(\frac{v}{1 + |\omega| v} \right)^{l + 1} \; , & \textrm{if } |\omega
v| \ll 1\\
\hspace*{-10mm} \textrm{and}  \\ 
 & | S_\omega (u,v)| &  \leq & \displaystyle C_3 \frac{v}{1 +|\omega| v }
e^{v|
\Im \omega|
 +u \Im \omega} \; , & \textrm{if } | \omega u| \gg 1.
\end{array} $$
The last inequality follows from the asymptotics 
$$|S_\omega(u,v)| \:  \sim \:
\Big|\frac{1}{\omega} \sin(\omega (u-v)) \Big| ,\quad
\textrm{if } | \omega u| \gg 1\; , $$ 
in the same way as the second inequality of Lemma \ref{lemma
zur abschaetzung fuer sinus}. Combining these cases we find a constant such that
\begin{equation} \label{eq: Abschaetzung fuer Greensfunktion}
| S_\omega (u,v) | \leq C_4 \left(\frac{u}{1 + |\omega| u} \right)^{-l}
\left(\frac{v}{1 + |\omega| v} \right)^{l + 1} e^{v| \Im \omega|
 +u \Im \omega} \; .
\end{equation}
Hence defining the function $Q_\omega$ by
\begin{equation} \label{eq: Definition Q_omega}
Q_\omega (u) := C_4 \int_u^\infty \frac{v}{1+ |\omega|v } \: |W_l(v)| \: dv \; ,
\end{equation}
which is well-defined for all $\omega \in F$ and $u\geq u_0 >0$ due to the
asymptotic of $W_l$,
it is straightforward to show inductively (cf. proof of Theorem \ref{theorem
ueber Loesungen der Jost equation}) that for all $m \in \N$
\begin{equation}  \label{eq: induktive Abschaetzung der einzelnen Summanden}
\big| \phi_\omega^{(m)}(u) \big| \leq C_1 \left( \frac{u}{1+ |\omega| u }
\right)^{-l} e^{u \Im \omega} \frac{Q_\omega (u)^{m}}{m!}\;.
\end{equation}
Now we proceed exactly as in the proof of Theorem \ref{theorem ueber Loesungen
der Jost equation}, where the inequality (\ref{eq: induktive Abschaetzung der
einzelnen Summanden}) can be considered as the analogue of (\ref{eq: induktive
Abschaetzung fuer iterationsschema}). It follows that the series (\ref{eq:
reihenansatz fuer integralgleichung im Besselfall})
converges locally uniformly in $\omega$ and $u$ and satisfies the integral
equation (\ref{eq: Integralgleichung fuer phi omega im besselfall}).
Furthermore, one shows inductively applying Lebesgue's dominated convergence
theorem, that for fixed $u$ each $\phi_\omega
^{(m)} (u)$ depends continuously of $\omega \in F$. It follows that the same 
is true for the series due to local uniform convergence.
\end{proof}

We are now ready to prove Theorem \ref{theorem: Konvergenz der Loesungen bei
omega = 0}:
\begin{proof}[Proof of Theorem \ref{theorem: Konvergenz der Loesungen bei
omega = 0}]
According to Lemma \ref{lemma: Konvergenz des Iterationsschemas}, our
perturbation
series (\ref{eq: reihenansatz fuer integralgleichung im Besselfall}) satisfies
the integral equation (\ref{eq: Integralgleichung fuer phi omega im
besselfall}). Using the recurrence formulas for the derivatives of $J_\nu(u)$,
one obtains
\begin{eqnarray*}
h_1'(l,\omega,u) & = & -\frac{l}{u} h_1(l,\omega,u) + \omega h_1(l-1, \omega,u)
, \\
h_2'(l, \omega,u) & = & -\frac{l}{u} h_2 (l,\omega,u) - \omega h_2(l-1,
\omega,u) \; , \textrm{ respectively.}
\end{eqnarray*}
This allow us to estimate the behavior of $\partial_u S_\omega (u,v)$. Exactly
as for $S_\omega (u,v)$, we obtain the following asymptotic formulas,
$$ |\partial_u S_\omega (u,v)| \leq C_5 \left(
\frac{u}{1 + |\omega| u}
\right)^{-l-1} \left( \frac{v}{1 + |\omega| v} \right)^{l+1} e^{v |\Im \omega|
+ u \Im \omega}\; .$$
Following the same arguments of the proofs of Theorems
\ref{Existenz der Loesungen acute phi grave phi} and \ref{theorem ueber
Loesungen der Jost equation}, and combining them with the above estimates and
asymptotic formulas we now have the following results:
\begin{enumerate} 
\item[1)] One can differentiate $\phi_\omega (u)$ with respect to $u$, and
$\phi_\omega '(u)$ is given by
\begin{equation} \label{eq: Ableitung von phi omega im Besselfall}
\phi_\omega '(u) =  \left( \phi_\omega^{(0)} \right) ' (u) - \int_u^\infty
\partial_u S_\omega (u,v) W_l(v) \phi_\omega (v) \: dv  \; . \nonumber
\end{equation}
In particular, Lebesgue's dominated convergence theorem yields that for fixed
$u$, $\phi_\omega '(u)$ is continuous in
$\omega \in F$.

\item[2)]
$\phi_\omega (u)$ and $\phi_\omega ' (u)$ obey the following estimates,
\end{enumerate}
\vspace{-5mm}
\begin{eqnarray*}
|\phi_\omega(u) - \phi_\omega^{(0)}(u)|  \leq  C_1 \left(
\frac{u}{1+|\omega|
u} \right)^{-l} e^{u \Im \omega} \left( e^{Q_\omega (u) }-1 \right)
\hspace{20mm} \\
\hspace{3mm} \big| \phi_\omega '(u) - \left( \phi_\omega ^{(0)} \right)'(u)
\big| \leq 
C_5 \left( \frac{u}{1+|\omega| u} \right)^{-l-1} e^{Q_\omega(u_0) } e^{u \Im
\omega} \int_u^\infty v|W_l(v)|dv \; .
\end{eqnarray*}
\begin{enumerate}
\item[]
Thus $\phi_\omega (u) \sim \omega^l e^{-i\omega u}$ and $\phi_\omega'(u) \sim
-i \omega^{l+1} e^{ -i \omega u}$ as $u \rightarrow \infty$.

\item[3)]
Differentiating $\phi_\omega (u)$ twice with respect to $u$ shows that
$\phi_\omega (u)$
is a solution of the Schr\"odinger equation (\ref{Schroedinger equation}) for
all $u \geq u_0$. Furthermore, from the asymptotics at infinity combined with
the uniqueness statement in Theorem \ref{Existenz der Loesungen acute phi grave
phi}, we know that
\begin{equation} \label{eq: phi omega ist omegahochl abstrich phi omega}
\phi_\omega (u) = \omega^l \grave{\phi}_\omega (u) \quad ,\textrm{ if }
\omega \neq 0, u\geq u_0 \; .
\end{equation}
Obviously, this extends to all $u \in \R$.
\end{enumerate}
Thus we have proven the continuity statement (\ref{eq: Stetigkeit
omegahochlphiomega gegen phi0}) for all $u\geq u_0$. On the other hand, we know
from the Picard-Lindel\"of theorem that
for $u$ on compact intervals, the solutions depend continuously on $\omega$.
This yields (\ref{eq: Stetigkeit omegahochlphiomega gegen phi0}) for all $u \in
\R$.

Finally, the asymptotics (\ref{eq: asymptotik der Loesung der SG fuer omega =
0}) is a simple consequence of (\ref{eq: anfangsfunktion im limes omega gegen
0}).
\end{proof}

\section{An Integral Spectral Representation}

In the previous section we derived some regularity results for the
solutions $\acute{\phi}_\omega$ and $\grave{\phi}_\omega$. We already know (cf.
Section \ref{section: construction of the resolvent}) that these solutions are a
system of fundamental solutions of the Schr\"odinger equation (\ref{Schroedinger
equation}) in the cases $\Im \omega < 0$ and $\Im \omega >0 $, respectively.
Thus the Wronskian $w(\acute{\phi}_\omega , \grave{\phi}_\omega )$ is
non-vanishing in these regions, which implies that the integral kernel
$k_\omega (u,v)$ of the resolvent is well defined. Since our next goal is to
get the limit in (\ref{eq: stones formula in unserem fall}), we prove
in the next lemma that the continuous extension of the solutions
$\acute{\phi}_\omega, \grave{\phi}_\omega$ to the real axis again yields a
system of fundamental solutions. More precisely,

\begin{lemma} \label{lemma: nichtverschwindende Wronski fuer omega reell}
The Wronskian $w(\acute{\phi}_\omega , \grave{\phi}_\omega )$ does not vanish
for $\omega \in \R \setminus \{0\}$. In particular, $\acute{\phi}_\omega,
\grave{\phi}_\omega$ are fundamental solutions for the Schr\"odinger equation
(\ref{Schroedinger equation}). In addition, this
remains true for the solutions $\acute{\phi}_0$ and $\phi_0$ in the case $\omega
= 0$.
\end{lemma}

\begin{proof}
Let us begin with the statement for $\acute{\phi}_0,\phi_0$: \\
For $\omega = 0$, the solutions $\acute{\phi}_0 (u) , \phi_0(u)$ have the
asymptotics
$$ \lim_{u \rightarrow - \infty} \acute{\phi}_0 (u) = 1 \quad \textrm{and} \quad
\lim_{u \rightarrow \infty} u^l \phi_0 (u) = (-i)^l (2l-1)!! \; .$$ 
Looking at the construction of these solutions, one sees that $\acute{\phi}_0$
is a real solution, while $\phi_0$ is either purely real or imaginary (depending
on the value of $l$). The Schr\"odinger equation for $\omega = 0$ reduces to
$\phi '' (u) = V_l(u) \phi (u)$ with a everywhere positive potential $V_l$.
Hence, exploiting the
special asymptotics, the solution $\acute{\phi}_0$ is convex and $\Re \phi_0$
($\Im \phi_0$, respectively) is either convex or concave depending on $l$.
In any case, we see that $\acute{\phi}_0$ and $\phi_0$ are linearly
independent, and thus $w (\acute{\phi}_0 ,\phi_0) \neq 0$. 

In order to prove the main part of the Lemma, we consider a complex solution $z
= z_1 +i z_2$ of the Schr\"odinger equation, where $\{ z_1, z_2 \}$ is a
fundamental system of real solutions, especially $w (z_1 ,z_2) \equiv c \neq
0$. Setting $y = \frac{z'}{z}$, a simple computation shows that
$$ \Im y  = \frac{w(z_1,z_2)}{|z|^2} \; , $$
where the right hand side is well defined because $w(z_1,z_2) \neq 0$
implies that $|z| \neq 0 $ everywhere. As a consequence, we have $\Im y \neq 0$
everywhere. Thus it follows that for all $u$ either $\Im y(u) > 0$ or $< 0$, due
to the continuity of the solution $z$ in $u$.

Applying this result to the solutions $\acute{\phi}_\omega$ and
$\grave{\phi}_\omega$,
respectively, and exploiting their asymptotics, one sees that $\Im
\acute{y}_\omega (u)$
and $\Im \grave{y}_\omega (u) $ have different signs for all $u$. Therefore,
$$ w(\acute{\phi}_\omega , \grave{\phi}_\omega ) = \acute{\phi}_\omega (u)
 \grave{\phi}_\omega '(u) - 
\acute{\phi}_\omega' (u) \grave{\phi}_\omega (u)= \acute{\phi}_\omega (u)
\grave{\phi}_\omega(u)  \big(\grave{y}_\omega (u) - \acute{y}_\omega (u)
\big) \neq 0 \; . $$
\end{proof}

As a consequence we have the following 

\begin{corollary}
The function $s_\omega (u,v)$ given by (\ref{s(u,v)}) is continuous in
 $ (\omega ,u ,v)$ for $\omega \in \{ \Im \omega \leq 0 \}$, $(u,v)
\in \R^2$.
\end{corollary}

\begin{proof}
We already know that for fixed $u_0 <0$, $\acute{\phi}_\omega(u_0) $ is
continuous in $\omega$ on $\{ \Im \omega \leq 0 \}$. Thus as solutions of the
linear differential equation (\ref{Schroedinger equation}), which depends
analytically on $\omega$ and smooth on $u$, the family $\acute{\phi}_\omega
(u)$ is (at least) continuous in $(\omega,u)$ in the region $\{ \Im \omega \leq
0\} \times \R $. Analogously this holds for $ \omega^l \grave{\phi}_\omega (u)$
according to
Theorems \ref{Existenz der Loesungen acute phi grave phi} and \ref{theorem:
Konvergenz der Loesungen bei omega = 0}. 
Since $s_\omega (u,v)$ is invariant if we substitute $ \omega^l
\grave{\phi}_\omega (u)$ for $\grave{\phi}_\omega (u)$, the
preceding lemma yields the claim.
\end{proof}

Note that the corollary is also true if $\omega$ is in the upper half
plane. The essential statement in this corollary is that one can extend
$s_\omega (u,v)$ continuously in $\omega$ up to the real axis.

>From the definitions (\ref{Defintion acute phi obere
Halbebene}) and (\ref{Definition grave phi obere Halbebene}), we have for
$\omega \in \{\Im \omega \neq 0 \}$  the relations 
$$ \overline{s_\omega (u,v)} = s_{\overline{\omega}} (u,v) \; , \quad
\textrm{hence} \quad \overline{k_\omega (u,v)} =  k_{\overline{\omega}} (u,v)
\;.$$
This allows us to simplify the expression (\ref{eq: stones formula in
unserem fall}). Evaluating for fixed $u$ the right hand side of (\ref{eq: stones
formula in unserem fall}) we obtain for any $\Psi \in
\algcal{H}$ as well as for any bounded interval $[a,b] \subseteq \R$
$$  \lim_{\epsilon \searrow 0} - \frac{1}{\pi} \int_a^b \left(\int_\R \Im
\big(k_{\omega - i \epsilon}(u,v) \big) \Psi (v) dv \right) d\omega \;.$$
According to the above corollary, we know that $\Im k_\omega (u,v) $ is
continuous in $(\omega,u,v)$ for $\omega \in \{ \Im \omega \leq 0\}$, $(u,v) \in
\R^2$. Thus, if we restrict $\Psi$ to the dense set $C^\infty _0(\R)^2$, we
integrate a continuous integrand over a compact interval. Hence, considering
the limit as a pointwise limit for any $u$, we may
interchange the limit and integration. Thus for any $\Psi \in
C^\infty_0
(\R)^2$, $[a,b] \subset \R$ bounded and $u$ the right hand side of (\ref{eq:
stones formula in unserem fall}) converges pointwise to
$$  - \frac{1}{\pi} \int_a^b\left( \int_{\textrm{supp}\: \psi} \Im \! \big(
k_\omega(u,v) \big) \psi(v) dv \right) d\omega \; . $$
Hence, together with the norm convergence in (\ref{eq: stones formula in
unserem fall}), the spectral projections of $H$ are for every $u$ described by
the formula
\begin{equation} \label{eq: spektralprojektoren fuer psi mit komp traeger}
\frac{1}{2} \big( P_{[a,b]} + P_{(a,b)} \big) \Psi (u) = - \frac{1}{\pi}
\int_a^b\left( \int_{\textrm{supp}\: \psi} \Im \! \big( k_\omega(u,v) \big)
\psi(v) dv \right) d\omega \; .
\end{equation}
In particular, this representation yields that $P_{[a,b]} \equiv P_{(a,b)}$.

\bigskip

As an immediate consequence we have the following 
\begin{corollary} \label{Cor: Spektrum von H ist absolutstetig}
The spectrum $\sigma (H)$ of the operator $H$ is absolutely continuous, i.e.
$\sigma(H) \equiv \sigma_{ac}(H)$.
\end{corollary}

\begin{proof}
The corollary is equivalent to the statement that the spectral measure $\langle
\Psi , dP_\omega \Psi \rangle$ of any $\Psi \in \algcal{H}$ is absolutely
continuous. 
This is clear by (\ref{eq: spektralprojektoren fuer psi mit komp traeger}) for
any $\Psi \in C_0^\infty (\R)^2$. But since this subset is dense, this also
holds on the whole Hilbert space $\algcal{H}$.
\end{proof}

\bigskip

Next we want to write the integrand in (\ref{eq:
spektralprojektoren fuer psi mit komp traeger}), i.e. $\int_{\mathrm{supp} \Psi}
... \: dv$, in a more compact way.
We first note that for real $\omega$ the complex conjugates of
$\acute{\phi}_\omega$ and $\grave{\phi}_\omega$ are again solutions of
(\ref{Schroedinger equation}). Hence, for any $\omega \in \R \setminus \{ 0\}
$ the pair $ \left\{ \acute{\phi}_\omega,\overline{\acute{\phi}_\omega}\right\}$
forms a fundamental system for this equation due to the boundary conditions.
Thus we can express $\grave{\phi}_\omega$ as a linear combination of
$\acute{\phi}_\omega$ and $\overline{\acute{\phi}_\omega}$,
$$ \grave{\phi}_\omega (u) =  \lambda(\omega) \acute{\phi}_\omega (u) + \mu
(\omega) \overline{\acute{\phi}_\omega (u)} \quad ( \omega \in  \R \setminus
\{0\}) \; ,$$ 
where $\lambda $ and $\mu$ are referred to as \textit{transmission
coefficients}.
The Wronskian of $\acute{\phi}_\omega$ and $\grave{\phi}_\omega$ can
be expressed by 
$$ w \left( \acute{\phi}_\omega, \grave{\phi}_\omega \right) =  \mu (\omega )
w \left( \acute{\phi}_\omega, \overline{\acute{\phi}_\omega} \right) =  -2 i
\omega \mu (\omega) \; ,$$
where in the last step we used the asymptotics (\ref{boundarycond3}). 
Moreover, we introduce the real fundamental solutions
$$ \phi_\omega ^1 (u) =  \Re \acute{\phi}_\omega (u) \; , \quad \phi_\omega^2
(u) = \Im \acute{\phi}_\omega (u)$$ 
and denote the corresponding eigenvectors of the Hamiltonian $H$ by \linebreak
$\Phi_\omega^a (u) = (\phi_\omega^a (u) , \omega \phi_\omega^a (u))^T$.

Using the above definitions, a short calculation shows that for $\omega \neq 0$
we can express the imaginary part of the Green's function $s_\omega (u,v)$ by 
\begin{equation} \label{eq: Neue Darstellung von Im s omega}
\Im s_\omega (u,v)= -\frac{1}{2\omega} \sum_{a,b = 1}^2 t_{ab}
(\omega) \phi_\omega^a (u) \phi_\omega^b(v) \; ,
\end{equation} 
where the coefficients $t_{ab} (\omega)$ are given by
\begin{eqnarray}
t_{11} (\omega) = 1 + \Re
\left(\frac{\lambda}{\mu} (\omega) \right) , & \quad t_{12}(\omega) = t_{21}
(\omega)
= - \displaystyle \Im \left( \frac{\lambda}{\mu}(\omega) \right) & \; ,
\nonumber\\
t_{22}(\omega) =  1 - \Re \left( \frac{\lambda}{\mu}(\omega) \right) \; . & &
\label{eq: Definition t_ab} 
\end{eqnarray} 
Since we know that $\Im s_\omega (u,v)$ is continuous for $\omega \in \R$ and
the expression (\ref{eq: Neue Darstellung von Im s omega}) holds for all
$ \omega \in \R \setminus \{0\}$, it extends to $\omega = 0$.
With (\ref{eq: Neue Darstellung von Im s omega}), the integrand in (\ref{eq:
spektralprojektoren fuer psi mit komp traeger}) can be written as
\begin{eqnarray*}
-\frac{1}{2\omega} \Big( \int\limits_{\mathrm{supp} \Psi} \!  \sum_{a,b = 1}^2
t_{ab} (\omega)
\phi_\omega^a (u) \phi_\omega^b(v) \left( \begin{matrix}
\omega & 1 \\ 
 \omega^2 & \omega 
\end{matrix} \right) \Psi (v) dv \Big) = \hspace{20mm}  \\
 = - \frac{1}{2\omega^2} \sum_{a,b=1}^2 t_{ab}(\omega)
\Phi_\omega^a (u) \Big( \int\limits_{\mathrm{supp} \Psi} \omega^2
\phi_\omega^b(v) \psi_1(v) + \omega \phi_\omega^b(v) \psi_2(v) \: dv
\Big) \; ,
\end{eqnarray*}
where the $\psi_i$ denote the two components of $\Psi$. 

Since $\phi_\omega^b(u)$ solves the Schr\"odinger equation
(\ref{Schroedinger equation}), it satisfies the relation \linebreak $
\big(-\partial_v^2 +V_l(v)\big)\phi_\omega^b(v) = \omega^2
\phi_\omega^b (v)$. Using this and integrating by parts, this simplifies to
\begin{equation} \label{eq: Darstellung des Integranden in
den Spektralprojektionen}
-\frac{1}{2\omega^2} \sum_{a,b=1}^2 t_{ab}(\omega) \Phi_\omega^a(u) \big\langle
\Phi_\omega^b, \Psi \big\rangle \; .
\end{equation}
(Note that in this case the energy scalar product of $\Phi_\omega^b$ and $\Psi$
is well defined, because $\Psi$ has compact support. Whereas in general this
does not
exist for arbitrary $\Psi \in \algcal{H}$, due to the fact that $\Phi_\omega^b
\notin \algcal{H}$.)

With (\ref{eq: Darstellung des Integranden in
den Spektralprojektionen}), we now obtain a more compact representation for
the spectral projections. Moreover, we can use (\ref{eq: Darstellung des
Integranden in den Spektralprojektionen}) to express the solution operators
$e^{-itH}$.

\begin{proposition} \label{lemma: abstrakte Darstellung der Loesungen, nicht
punktweise}
Consider the Cauchy Problem (\ref{Hamiltonform}) for compactly supported smooth
initial data
$\Psi_0 \in C^\infty_0 (\R)^2$. Then the solution has the integral
representation
\begin{eqnarray}
\Psi(t) & = & e^{-itH} \Psi_0  = \nonumber \\
& = & \frac{1}{2\pi} \int_\R e^{-i \omega t} \frac{1}{\omega^2} \sum_{a,b=1}^2
t_{ab}(\omega) \Phi_{\omega }^a \: \big \langle \Phi_{\omega }^b , \Psi_0
\big \rangle \: d\omega \; . 
\label{eq: Integraldarstellung der Loesung: abstrakt}
\end{eqnarray}
Here the integral converges in norm in the Hilbert space $\algcal{H}$.
\end{proposition}

\begin{proof}
We use the following variation of Stone's formula to obtain for any
bounded interval $(c,d) \subseteq \R$
\begin{eqnarray*}
\frac{1}{2}  e^{-itH} \left( P_{[c,d]} + P_{(c,d)}\right) \Psi \hspace{60mm} \\
= \lim_{\epsilon \searrow 0} \int_c^d e^{-i\omega t} \left[(H- \omega -i
\epsilon)^{-1} - (H-\omega + i\epsilon)^{-1} \right] \Psi \:d\omega \; ,
\end{eqnarray*} 
where the limit is with respect to the norm of $\algcal{H}$.
Since we know that \linebreak $P_{[c,d]} \equiv P_{(c,d)}$, it follows that
this expression is equal to $e^{-itH} P_{(c,d)} \Psi$. Using the explicit
formula for the resolvent, for every $u\in \R$ the right hand side is equal to
\begin{eqnarray}   \lim_{\epsilon \searrow 0}  -\frac{1}{\pi} \int_c^d
e^{-i\omega t} \left(\int_\R \Im \! \big(k_{\omega-i \epsilon}(u,v)\big)
\Psi(v) dv \right) d\omega \; . \label{eq: zwischenschritt im beweis der 
proposition zur abstrakten integraldarstellung}
\end{eqnarray}
Due to the continuity of the imaginary part of the kernel $k_\omega(u,v)$, we
may take for $\Psi_0 \in C^\infty_0(\R)^2$ and $(c,d)$ bounded the
pointwise limit for any $u \in \R$. Hence, using (\ref{eq: Darstellung des
Integranden in den Spektralprojektionen}) we can simplify (\ref{eq:
zwischenschritt im beweis der proposition zur abstrakten integraldarstellung})
to
$$ \frac{1}{2\pi} \int_c^d e^{-i \omega t}
\frac{1}{\omega^2} \sum_{a,b=1}^2 t_{ab}(\omega) \Phi_{\omega }^a (u) \big
\langle \Phi_{\omega }^b , \Psi_0
\big \rangle \: d\omega \; , $$
and together with the norm convergence it follows that this term is equal to
$e^{-itH} P_{(c,d)} \Psi_0 (u)$.
Using the abstract spectral theorem and that $e^{-itH}$ is a unitary operator,
it is clear that $e^{-itH}P_{(-n,n)}\Psi_0 \rightarrow e^{-itH}\Psi_0$ in norm
as $n \rightarrow \infty$.
\end{proof}

This proposition extends to the following theorem.
\begin{theorem} \label{theorem: punktweise Darstellung der Loesungen}
For any fixed $u \in \R$ the integrand in the representation
(\ref{eq: Integraldarstellung der Loesung: abstrakt}) is in $L^1(\R, \C^2)$ as
a function of $\omega$. 
In particular, the representation (\ref{eq: Integraldarstellung der Loesung:
abstrakt}) of the solutions holds pointwise for every $u\in \R$, i.e.
\begin{equation} \label{eq: punktweise integraldarstellung der Loesung}
\Psi(t,u) = \frac{1}{2\pi} \int_\R e^{-i \omega t} \frac{1}{\omega^2}
\sum_{a,b=1}^2 t_{ab}(\omega) \Phi_{\omega }^a (u) \: \big \langle \Phi_{\omega
}^b , \Psi_0 \big \rangle \: d\omega \; .
\end{equation}
Moreover, for $u$ fixed, the function $\Psi (t,u)$ vanishes as $t \rightarrow
\infty$.
\end{theorem}

\begin{proof}
Since we know that the integrand is continuous in $\omega$, it is in
$L^1([a,b],\C^2)$ for any bounded interval $[a,b]$. Thus it remains to
analyze the integrand for large $|\omega|$.

To this end, we must investigate the asymptotic behavior of the
fundamental solutions $\acute{\phi}_\omega$ and $\grave{\phi}_\omega$ in
$\omega$. We constructed these solutions with the
iteration scheme (\ref{Iterationsschema fuer phi 1}) as solutions of the Jost
equation. For the proof of this, the estimate (\ref{eq: Schranke fuer G_omega})
played an essential role. Since in this case we consider real $\omega$ with
$|\omega| \gg 1$, we can use the simple estimate
$\big|\frac{1}{\omega} \sin(\omega(u-v))\big| \leq \frac{1}{| \omega|}$ instead
of (\ref{eq: Schranke fuer G_omega}), which now holds for every $u,v \in \R$.
Thus, proceeding exactly in the same way as in the proof of Theorem \ref{theorem
ueber Loesungen der Jost equation}, we now obtain the following estimates for
the several terms in the series expansion (\ref{Reihenentwicklung von phi 1})
$$ \big| \phi_\omega^{(k)} (u) \big| \leq \frac{1}{k !} \hat{P}_\omega (u) ^k \;
, \quad \textrm{where } \hat{P}_\omega (u) := \int_{-\infty}^u
\frac{1}{|\omega|} V_l(v) dv \; ,$$ 
for any $k \in \N$ and $u \in \R$. Thus the
solution $\acute{\phi}_\omega (u)$ ($\omega \neq 0$) is given \textit{for all}
$u \in \R$ by this series expansion and obeys the uniform bound $$\big|
\acute{\phi}_\omega (u) - e^{i \omega u} \big| \: \leq \: e^{\hat{P}_\omega (u)}
- 1 \: \leq \: e^{\frac{1}{|\omega|} \Vert V_l \Vert_{L^1}} -1\; ,$$ 
since $V_l \in L^1(\R)$.
In particular,
\begin{equation} \label{eq: Abschaetzung fuer acute phi, omega reell}
 \big| \acute{\phi}_\omega (u) \big| \leq 1 +
\mathcal{O}\left(\frac{1}{|\omega|}\right) \quad \textrm{for } |\omega|\gg 1 \;
. \end{equation}
Next we investigate the $\omega$-dependence of $\big\langle\Phi_\omega^b,\Psi_0
\big\rangle$. We integrate by parts to obtain 
$$ \big\langle \Phi_\omega^b, \Psi_0  \big \rangle = \int
\limits_{\mathrm{supp} \Psi_0} \phi_\omega^b(v) \big( \omega \psi_2(v) -
\psi_1(v)''
+ V_l(v) \psi_1(v)\big) dv \; ,$$ where $\Psi_0 = (\psi_1,\psi_2)^T$ (note that
the
boundary terms drop out, because $\Psi_0 \in C^\infty_0(\R)^2$). Since
$\phi_\omega^b(u)$ is a solution of the Schr\"odinger equation
(\ref{Schroedinger
equation}), we substitute $\frac{1}{\omega^2} \left( - \phi_\omega^b(u) '' +
V_l(u) \phi_\omega^b(u)\right)$ for $\phi_\omega^b (u)$ and integrate by
parts twice,
$$ = \frac{1}{\omega^2} \int \limits_{\mathrm{supp} \Psi_0} \phi_\omega^b
\big( - (\omega \psi_2 - \psi_1 '' + V_l \psi_1 )'' + V_l (\omega \psi_2 -
\psi_1'' + V_l \psi_1) \big) dv \; .$$
We can now iterate this procedure as often as we like due to the fact that
$\Psi_0 \in C^\infty_0(\R)^2$ and $V_l \in C^\infty (\R)$. Thus using the bound
(\ref{eq: Abschaetzung fuer acute phi, omega reell}), we obtain arbitrary
polynomial decay in $\omega$ for $\big \langle \Phi_\omega^b , \Psi_0 \big
\rangle$. 

Thus it remains to control the coefficients $t_{ab}(\omega)$ for large
$|\omega|$. According to the definition of the transmission coefficients
$\lambda
(\omega)$ and $\mu(\omega)$, they satisfy the following relations,
$$w( \grave{\phi}_\omega, \acute{\phi}_\omega ) =  2i \mu (\omega) \quad
\textrm{and} \quad w(\grave{\phi}_\omega, \overline{\acute{\phi}_\omega}) =  2i
\lambda (\omega)\; .$$
In order to calculate the Wronskians, we substitute the Jost integral
equations (\ref{Jost equation bound cond -unendlich}),(\ref{eq: Jost gleichung
randbedingungen +unendlich}) for $\acute{\phi}_\omega$ and
$\grave{\phi}_\omega$, respectively, as well as the corresponding integral
equations for the derivatives (for instance (\ref{eq: Ableitung von phi omega})
in the case $\acute{\phi}_\omega$) and obtain immediately
$$ \mu(\omega) =  1 + \mathcal{O} \left(\frac{1}{\omega}\right)\; , \quad
\lambda(\omega) = \mathcal{O} \left(\frac{1}{\omega}\right).$$
Hence the coefficients $t_{ab}(\omega)$ remain (at least) bounded, according
to their definition (\ref{eq: Definition t_ab}).

We conclude that the integrand in (\ref{eq: punktweise integraldarstellung der
Loesung}) is in $ L^1(\R ,\C^2)$ as a function of $\omega$. Thus
the right hand side in the integral representation (\ref{eq: Integraldarstellung
der Loesung: abstrakt}) converges also pointwise and, together with the norm
convergence, (\ref{eq: punktweise integraldarstellung der Loesung}) follows.

Since for $u$ fixed, $\Psi (t,u)$ is a Fourier transform of a $L^1$ -function,
the \linebreak Riemann-Lebesgue lemma applies. Hence $\Psi(t,u)$ vanishes as $t
\rightarrow \infty$.
\end{proof}

In the next step we extend this proposition to the Cauchy problem (\ref{eq:
Cauchy problem in 4 Dimensionen}).

\begin{theorem} \label{theorem: Darstellung der Loesung des Cauchy problems in
4dim}
Consider the Cauchy problem (\ref{eq: Cauchy problem in 4 Dimensionen}) for
smooth and compactly supported initial data. Then there exists a unique smooth
solution, which is compactly supported for all times.

Moreover, decomposing the initial data $\Psi_0$ into spherical harmonics, the
solution has the representation
\begin{equation} \label{eq: Darstellung der Loesung des Cauchy problems in 4dim}
\Psi(t,u,\vartheta,\varphi) =  \sum_{l=0}^\infty \sum_{|m| \leq l} e^{-itH_l}
\Psi_0^{lm} (u) Y_{lm}(\vartheta, \varphi) \; .
\end{equation}
\end{theorem}
\begin{proof}
For the existence and uniqueness of such a solution we apply the theory of
linear symmetric hyperbolic systems (cf. \cite{John}). Since the equation in
(\ref{eq: Cauchy problem in 4 Dimensionen}) is defined on $\R \times \R \times
S^2$ we have to work in local coordinates for $S^2$. We demonstrate the idea
in the chart $\big(U,(\vartheta,\varphi)\big)$, where $U$ is an open, relative
compact subset of $S^2$ such that $(\vartheta, \varphi)$ are well defined on
$\overline{U}$.

Letting $\Phi = (\partial_t \psi, \partial_u \psi, \partial_{\cos \vartheta}
\psi, \partial_\varphi \psi, \psi)^T$ we can write the equation as the first
order system
$$ A^0 \partial_t \Phi + A^1 \partial_u \Phi + A^2 \partial_{\cos \vartheta}
\Phi +A^3 \partial_\varphi \Phi + B \Phi = 0 \; , $$ 
where the matrices $A^i,B$ are given by
$$ A^0 = \mathrm{diag}\left(1,1,\left(1-\frac{2M}{r}\right) \frac{1}{r^2} \sin^2
,\left(1-\frac{2M}{r}\right) \frac{1}{r^2} \frac{1}{\sin^2 \vartheta},1\right)
\; ,$$
$$A^1 = (a_{ij}^1) \; , \quad \textrm{with} \quad 
a_{12}^1 = a_{21}^1 = -1   \; ,\hspace{30mm}$$
$$A^2 = (a_{ij}^2) \; , \quad \textrm{with} \quad 
a_{13}^2 = a_{31}^2 = - \left(1-\frac{2M}{r}\right) \frac{1}{r^2} \sin^2
\vartheta \; ,$$
$$ A^3 = (a_{ij}^3) \; , \quad \textrm{with} \quad  
a_{14}^3 = a_{41}^3 = - \left(1-\frac{2M}{r}\right) \frac{1}{r^2}
\frac{1}{\sin^2 \vartheta}  \; ,$$
\begin{eqnarray*}
 B = (b_{ij}) \; , & \textrm{ with }\hspace{2mm} &  
b_{13} = \left(1 - \frac{2M}{r}\right) \frac{1}{r^2} 2 \cos \vartheta \; ,  \\ 
 & & b_{15} = \left(1 -\frac{2M}{r}\right) \frac{2M}{r^3}, \; b_{51} = -1
 \; ,
\end{eqnarray*}
and all other coefficients vanish. By multiplying this system with $ \left( 1-
\frac{2M}{r}\right)^{-1}r^2$, we obtain a linear symmetric hyperbolic
system on
$\R \times \R \times U$ in the sense that the $A^i$ are symmetric and $A^0$ is
uniformly positive definite on $\R \times \R \times U$. Since the initial data
$\Psi_0$ has compact support, we can restrict the system to $\R \times V \times
U$,
where $V \subseteq \R$ is an open, relative compact set with $ \mathrm{supp}
\Psi_0
\subseteq V$. Considering the system on this domain, the matrices $A^i,B$ remain
uniformly bounded. Since we can cover $S^2$ by a finite number of such
charts, the theory of symmetric hyperbolic systems yields an $\epsilon_1
>0 $ such that there is unique and smooth solution $\psi(t,u,x)$ for all $t
<\epsilon_1$ on $\R \times V \times S^2$ with initial data $\Psi_0$.

Moreover, this solution has finite propagation speed, which is independent of
$u$ (this is physically clear from causality; it follows mathematically by
considering lens-shaped regions for our symmetric hyperbolic systems).
Thus there exists an $\epsilon >0$ (possibly smaller than $\epsilon_1$) such
that the solution $\psi(t,u,x)$ has compact support in $V \times S^2$ for all
times $t \leq \epsilon$. Iterating this argument for the Cauchy problem with
initial data $(\psi(\epsilon,u,x), i \partial_t \psi(\epsilon,u,x))$ (and
choosing $V \subseteq \R$ sufficiently large), we get a unique and smooth
solution $\psi(t,u,x)$ with compact support for all $t \leq 2 \epsilon$ and so
forth.
Thus we have proven the existence of a global solution $\psi(t,u,x) \in
C^\infty(\R \times \R \times S^2)$ which is unique and compactly supported for
all times $t$.

In order to prove the representation (\ref{eq: Darstellung der Loesung des
Cauchy problems in 4dim}), we consider the restriction of the solution $\Psi
(t,u,x) = (\psi(t,u,x) , i\partial_t \psi(t,u,x))^T$ of the Cauchy problem
(\ref{Hamiltonform allgemein}) in Hamiltonian form to fixed modes $l,m$
$$ \Psi^{lm}(t,u) Y_{lm}(\vartheta,\varphi) = \langle \Psi(t,u), Y_{lm}
\rangle_{L^2(S^2)} Y_{lm}(\vartheta, \varphi) \; .$$
Then $\Psi^{lm}(t,u) Y_{lm}(\vartheta,\varphi)$ is a solution of
(\ref{Hamiltonform allgemein}) with initial data $\Psi_0^{lm}(u)
Y_{lm}(\vartheta,\varphi)$, which is smooth and compactly supported. Thus
$\Psi^{lm}(t,u)$ is a solution of the Cauchy problem (\ref{Hamiltonform}), and
due to the uniqueness of such solutions 
$$\Psi^{lm} (t,u) = e^{-itH_l} \Psi_0^{lm} (u)\; .$$
Now the uniqueness of the decomposition into spherical harmonics yields
(\ref{eq: Darstellung der Loesung des Cauchy problems in 4dim}).
\end{proof}

We are now ready to prove our main theorem.

\begin{proof} [Proof of Theorem \ref{theorem: Haupttheorem}]
The existence and uniqueness of solutions of the Cauchy problem follow
directly from Theorem \ref{theorem: Darstellung der Loesung des Cauchy problems
in 4dim} after the substitution $\phi = r \psi$. Thus it remains to show the
pointwise decay.

The conserved energy for solutions which are compactly supported for all
times $t$ implies that for every $t$
$$ \Vert \Psi(t,u,\vartheta,\varphi)\Vert^2 = \Vert \Psi_0
(u,\vartheta,\varphi) \Vert^2 = \sum_{l=0}^\infty \sum_{|m| \leq l} \Vert
\Psi_0^{lm} (u) \Vert_l^2 \; ,$$
where for the second equation we used the isometry (\ref{eq: Isometrie bzgl der
Zerlegung in Moden}). Hence, defining 
$$ \Psi^{L} (t,u,\vartheta, \varphi) := \sum_{l=L}^{\infty} \sum_{|m| \leq l}
\Psi^{lm}(t,u) Y_{lm}(\vartheta,\varphi) \; , $$ 
we can find for every $\epsilon > 0$ a number $L_0$ such that
$$\Vert \Psi^{L_0} (t,u,\vartheta, \varphi) \Vert^2 = \sum_{l= L_0}^\infty
\sum_{|m| \leq l} \Vert \Psi_0^{lm}(u) \Vert_l^2 < \epsilon \; .$$
Let us now consider the Cauchy problem (\ref{eq: Cauchy problem in 4
Dimensionen}) with initial data $$H \Psi_0 = \sum_{l=0}^\infty \sum_{|m|\leq l}
(H_l \Psi_0^{lm} )Y_{lm} \; .$$ 
Obviously, this data is also smooth and compactly supported and thus gives rise
to
the solution
$$ \sum_{l=0}^\infty \sum_{|m|\leq l} \left(e^{-itH_l} H_l \Psi_0^{lm}\right)
Y_{lm} =  \sum_{l=0}^\infty \sum_{|m|\leq l} \left(H_l e^{-itH_l} \Psi_0^{lm}
\right) Y_{lm} = H \Psi \; ,$$
where in the second equation we again used the uniqueness of the decomposition
into spherical harmonics. Thus for every $\epsilon > 0$ there is a $L_1$
(without restriction $\geq L_0$) such that 
$$ \Vert H \Psi^{L_1} (t) \Vert < \epsilon \; , \quad \textrm{for all times } t
\; .$$
Proceeding inductively, we find for every number $N$ and for every $\epsilon
>0$ a number $L_N$ such that
$$\Vert H^n \Psi^{L_N} (t) \Vert < \epsilon \; , \quad \textrm{for all } t
\textrm{ and } n\leq N \; .$$ 
Let $K \subseteq \R \times S^2$ be an arbitrary compact subset with smooth
boundary. Then, due to the
definition of the energy, there exists a constant $C_0(K) >0$ such that for
$\Psi^{L_N} = (\psi_1^{L_N} , \psi_2^{L_N})^T$,
$$ \Vert \psi_1 ^{L_N} \Vert_{H^1(K)} + \Vert \psi_2^{L_N} \Vert_{L^2(K)} \:
\leq \: C_0(K) \Vert \Psi^{L_N} \Vert \; .$$
Applying the same argument to $H \Psi^{L_N}  = (\psi_2^{L_N} , A
\psi_1^{L_N})^T$, where $A$ is the differential operator given by (\ref{eq:
Differentialoperator A}), there is a $C_1(K) >0$ such that
$$ \Vert A \psi_1^{L_N} \Vert_{L^2(K)} + \Vert \psi_2^{L_N} \Vert_{H^1(K)} \:
\leq \: C_1(K) \Vert H \Psi^{L_N} \Vert \; .$$
Since the differential operator $A$ is of the form $A = -\Delta + X$, where
$X$ is a first order differential operator, it is in particular a second
order elliptic partial differential operator. Thus, for $u \in C^\infty(\R
\times S^2)$ and for each $U \subset \subset V \subset \subset \R \times S^2$
($\subset \subset$ denotes relative compact) there is an estimate (cf.
\cite[p.379 (11.3)]{Taylor1})
$$ \Vert u\Vert_{H^{k+2}(U)} \leq C \Vert A u\Vert_{H^k(V)} + C \Vert u \Vert_
{H^{k+1}(V)} \quad \textrm{for all } k\geq 0 .$$
It follows that there exist new constants $C_0(K), C_1(K)$ such
that
$$\Vert \psi_1^{L_N} \Vert_{H^2(K)} + \Vert \psi_2^{L_N}\Vert_{H^1(K)} \: \leq
\:  C_0 (K) \Vert \Psi^{L_N} \Vert + C_1 (K) \Vert H \Psi^{L_N}
\Vert \; .$$
Iterating this inequality, we obtain constants $C_0(K),...,C_k(K)$ such that
$$ \Vert \psi_1^{L_N} \Vert_{H^{k+1}(K)} + \Vert \psi_2^{L_N}\Vert_{H^k(K)} \:
\leq
\:  \sum_{n=0}^k C_n(K) \Vert H^n \Psi^{L_N}
\Vert \; .$$
In particular, for every $\epsilon >0$ there is a number $L$ such that
$$\Vert \Psi^L(t) \Vert_{H^2(K)} < \epsilon \quad \textrm{for all } t \;
.$$
Thus the Sobolev embedding theorem yields (possibly after enlarging $L$)
$$\Vert \Psi^L(t) \Vert_{L^\infty(K)} < \epsilon \quad \textrm{for all } t \;
.$$
Furthermore, due to the pointwise decay for fixed modes $l,m$ which was shown in
Theorem \ref{theorem: punktweise Darstellung der Loesungen}, we can find for any
$\epsilon > 0$ and $(u,x) \in \R \times S^2$ a time $t_0$ and a number $L$
such that for the solution $\Psi(t,u,x)$ of the Cauchy problem (\ref{eq: Cauchy
problem in 4 Dimensionen}),
$$ | \Psi (t,u,x)| \leq \sum_{l=0}^{L-1} \sum_{|m| \leq l} | \Psi^{lm}(t,u)
Y_{lm}(x)| + |\Psi^L (t,u,x)| < \epsilon \quad \textrm{for all }t\geq t_0 \; .$$
Since $\phi = r \psi$, this concludes the proof.
\end{proof}

\bigskip

\begin{acknowledgment}
The author would like to thank Felix Finster, University of Regensburg, for
introducing to this problem and also for his helpful suggestions and remarks.
\end{acknowledgment}

\noindent
NWF I -- Mathematik,
Universit{\"a}t Regensburg, 93040 Regensburg, Germany, \\
{\tt{Johann.Kronthaler@mathematik.uni-regensburg.de}}

\end{document}